\documentclass[twocolumn]{autart}

\usepackage{amsmath}
\usepackage{amssymb}
\usepackage{amstext}
\usepackage{amsbsy}
\usepackage{amsopn}
\usepackage{epsfig}
\usepackage{theorem}
\usepackage{fancybox}
\usepackage{shadow}
\usepackage{float}
\usepackage{bm}
\usepackage{graphicx}
\usepackage{pstool}
\usepackage{tikz}
\usepackage{amsfonts}

\usepackage{latexsym}

\usepackage{comment}
\usepackage{color}
\usepackage{enumerate}
\usepackage{psfrag}
\usepackage{empheq}
\usepackage{cite}
\usepackage{epstopdf}

\def \RR {{\mathbb R}}

\def \bmu {\bm u}
\def \barbmu {\bar{\bm u}}
\def \barbarbmu {\bar{\bar{\bm u}}}
\def \bmp {\bm p}
\def \bmq {\bm q}
\def \bmtildep {{\bm \tilde{\bm p}}}
\def \bmtildepF {{\bm \tilde{\bm p}}_{\mathcal{F}}}
\def \dotbmtildepF {\dot{\bm \tilde{\bm p}}_{\mathcal{F}}}

\def \dotbmtildep {\dot{\bm \tilde{\bm p}}}
\def \Piu {\Pi_{\bmu}}
\def \sat {\mbox{sat}}

\newtheorem{propo}{Proposition}
\newtheorem{lemma}{Lemma}
\newtheorem{theo}{Theorem}
\begin{document}
\begin{frontmatter}
\title{A unified approach to\\ fixed-wing aircraft path following guidance and 
control}
\author[first]{J.-M. Kai}
\author[Third]{T. Hamel}
\author[Fifth]{C. Samson}
\address[first]{I3S, Universit\'e C\^ote d'Azur, CNRS, Sophia Antipolis, France \\(e-mail: $kai@i3s.unice.fr$)}
\address[Third]{I3S, Universit\'e C\^ote d'Azur, Institut Universitaire de France, CNRS, Sophia Antipolis, France \\(e-mail: $thamel@i3s.unice.fr$)}
\address[Fifth]{C. Samson (corresponding author) is with INRIA and I3S UCA-CNRS, Sophia Antipolis, France, (email: $claude.samson@inria.fr$, $csamson@i3s.unice.fr$).}




\vspace{-0.2cm}
\begin{abstract}
This paper addresses the path following control problem for scale-model fixed-wing aircraft. Kinematic guidance and dynamic control laws are developed within a single coherent framework that exploits a simple generic model of aerodynamic forces acting on the aircraft and applies to almost all regular 3D paths. The proposed control solutions are derived on the basis of theoretical stability and convergence analyses. They are complemented by addressing several practical issues, and validated via realistic hardware-in-the-loop simulations.
\end{abstract}

\vspace{-0.2cm}
\end{frontmatter}
\section{Introduction}
\vspace{-0.2cm}
During the last two decades important progress has been made in the development of advanced methods for motion control of autonomous Unmanned Aerial Vehicles (UAVs). The commercial landscape of these vehicles is characterized by a plethora of small start-up companies proposing an increasing variety of devices for various applications such as power line inspection, data collection, crop monitoring, real estate use of aerial photographs, etc. This rapidly growing market has in turn boosted research studies in the field of UAVs, particularly from the Systems and Control community, with specific control strategies developed in relation to different classes of aerial vehicles, as exemplified by scale-model airplanes, quadrotors and, more recently, tail-sitters.
\newline A typical aerial vehicle is an underactuated dynamic system with six main degrees of freedom (for the position and orientation of the vehicle's main body) and usually four (sometimes three) independent control inputs, i.e. a thrust force along the main direction of the vehicle, and three (or only two) control torques for body orientation modification. Associated motion control problems are commonly classified into two sub-categories, namely {\em trajectory tracking} and {\em path following}.
\newline Trajectory tracking refers to the problem of stabilizing a time-parametrized reference position trajectory. It is best adapted to highly maneuverable Vertical-Take-Off-and-Landing (VTOL) vehicles (quadrotors, helicopters, ducted-fans) \cite{mellinger2011,lee2010,hua2013introduction,pucci2015nonlinear}, convertible vehicles (tail-sitters in particular) \cite{zhou2017} and acrobatic aircraft \cite{sonneveldt2009nonlinear} whose positions have to be precisely and timely monitored near hovering. Its applicability to fixed-wing aircraft, although possible \cite{ren2004trajectory,KaiHamelSamsonCDC2017}, is more limited because the position-timing issue for cruising vehicles is less essential than following a preplanned path with a given (possibly high) velocity needed for the production of strong air-lift forces on profiled wings that dramatically reduce energy expenditure.

Path-following refers to the problem of stabilizing a desired forward velocity and of zeroing the distance between the vehicle position and a desired geometric path \cite{samson1992path}. 
The path itself may be attached to a frame that is moving with respect to some fixed inertial frame \cite{OAE2016}. Path following control has become a central feature of any advanced automated flight control system. It can be used for waypoint navigation, to fly a programmed course \cite{nelson2007vector},
and for a landing approach with a given glide slope \cite{Lebras2014}. It is commonly decomposed into sub-problems, namely guidance (sometimes termed as guidance-based control), speed control, and attitude control. Guidance essentially consists in determining the desired heading direction for the aircraft given a path to follow \cite{BreivikFossen2005}, and involves body kinematic equations only. It is mostly independent of the aircraft characteristics and is usually associated with the notion of outer-loop control. By contrast, velocity and attitude controls take into account the specificities of the aircraft and of aerodynamic reaction forces that influence its dynamics. These lower level control tasks are associated with the notion of inner-loop control. Automatic control of the aircraft speed is commonly done via thrust production and is sometimes called auto-throttle control. Automatic control of the aircraft attitude (or orientation) is in charge, via the production of torques, of making the actual aircraft heading direction converge to the direction specified at the guidance level and is typically performed by modifying the orientation angles of the aircraft control surfaces. For this latter task three-axis autopilots monitor angles of the aircraft ailerons (for roll motion), elevator(s) (for pitch motion), and rudder(s) (for yaw motion). Less sophisticated two-axis autopilots monitor roll and pitch motions only. The aforementioned separation between kinematic guidance and dynamic speed and attitude control is conceptually attractive and convenient, all the more so that imprecise knowledge of aerodynamic forces acting on the aircraft dynamics constitutes the main source of difficulty for the design of robust controllers. Inner-loop controllers have historically, and to these days, been essentially designed on the basis of linearized dynamic equations about so-called {\em trim} trajectories (for which the aircraft longitudinal and angular velocities are constant in the absence of wind) with attack and bank angles often taken as intermediary control variables \cite{BrianL92,Beard:2012}. They are also reported in all major flight dynamics textbooks and, for this reason, are often taken for granted in path following studies that focus only on the simpler generic guidance part of the problem. Among the most advanced ones let us cite, for instance, those based on linear control techniques combined with gain scheduling \cite{Carter96}, feedback linearisation combined with LQ control \cite{Escande97}, or Linear Parameter Varying (LPV) modeling and control \cite{LPV_aircraft,CAI2011}.

The central contribution of the present paper is the development of a nonlinear control approach that encompasses within a single framework the different steps involved in aircraft path following control, i.e. guidance, speed control and attitude control. This approach comes as a complement to our previous work devoted to the trajectory tracking problem \cite{KaiHamelSamsonCDC2017,pucci2015nonlinear}. It offers an alternative to current state-of-the-art automated flight solutions, and goes further in terms of convergence and stability analyses over an extended flight domain. For instance, the desired path is not required to be a trim trajectory, and convergence can be achieved whatever the initial distance that separates the aircraft from the path.\\ 
The proposed control design methodology exploits a simple model of aerodynamic forces applied to the vehicle that is both representative of the physics underlying the creation of the environmental forces and sufficiently simple to lend itself to control design and analysis. The attentive reader will note that this model assumes very little about the sideslip transversal component of this force. This is coherent with the difficulty of working out a good model of this component. Nor does it limit the attack angle, by contrast with most commonly used models.
The proposed control laws are complemented with bounded integral actions to compensate for inevitable modeling errors, and their properties are established assuming a perfect modeling of the aerodynamic forces in the plane orthogonal to the main wing. Omitting integral terms would have facilitated the control design and associated analyses, but this would have been at the expense of good practice.\\   
\newline The paper is organized as follows. Section \ref{model} recalls the equations of motion of an aircraft, presents the model used for the aerodynamic forces acting on an aircraft, and specifies the control variables involved in a {\em generic} control design that does not depend on specific aircraft actuation means. Section \ref{control} first recalls the general control objectives associated with the path following problem and some useful relations related to the calculation of the distance between the aircraft and the desired path. The control design itself involves three stages: i) thrust control of the aircraft inertial speed, ii) design of a kinematic guidance vector yielding convergence to the desired path with the pre-specified route direction, and iii) determination of a desired body frame whose asymptotic stabilization yields a balanced flight, i.e. sideslip angle zeroing, and convergence of the aircraft heading vector to the previously defined guidance vector. These three stages are subsequently detailed, yielding control laws whose convergence and stability properties are summarized in four propositions and two theorems. For the sake of not hindering the reading, the proofs of these propositions and theorems are reported in appendices. The two assumptions under which local exponential stability is established are then particularized and discussed in the common cases of straight and circular paths. Section \ref{complementary} addresses complementary practical issues, namely i) extension of the control design to curves defined w.r.t. a translating frame, ii) airspeed control instead of inertial speed control, iii) adaptation to two-axis autopilots, iv) monitoring of thrust limitations and attack angle for stall avoidance, and v) determination of control surfaces angles from desired aircraft angular velocity. Hardware-in-the-loop simulation results involving a scale-model aircraft and challenging reference paths, with large initial position errors, imposed positive thrust, unknown wind and air-velocity measurement approximations, purposely introduced to test the control performance and its robustness, are reported in Section \ref{simulations}. Concluding remarks
are given in the last Section \ref{conclusion}.

\section{Control model}\label{model}
\vspace{-0.2cm}

\subsection{Notation} \label{sec-back}
\begin{figure}
 \includegraphics[width=.47\textwidth]{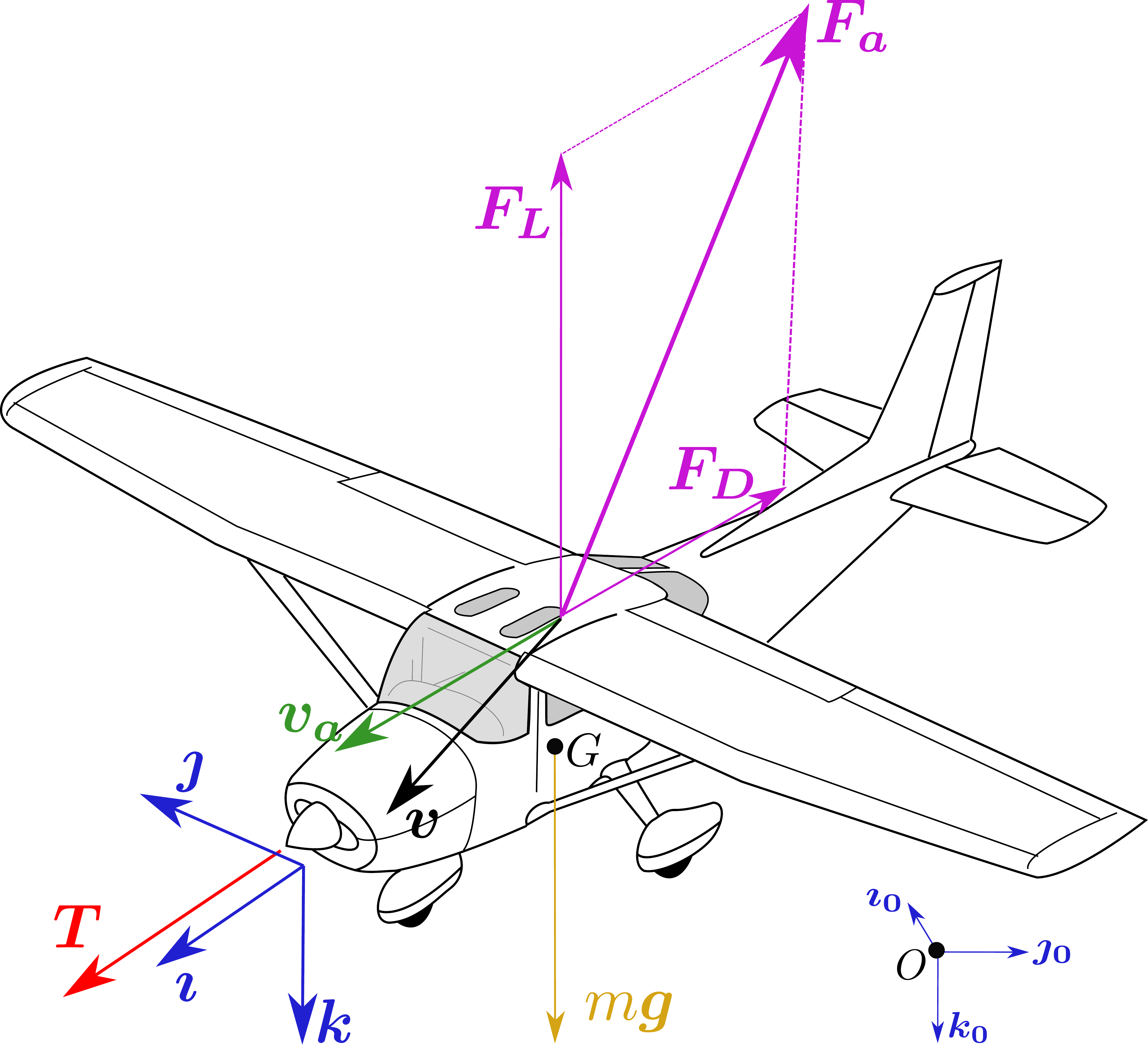}
 \caption{Frames and forces}
 \label{frames}
 \label{planemodel}
\end{figure}

\vspace{-0.2cm}
Throughout the paper, ${\bm E}^3$ denotes the $3D$ Euclidean vector space and vectors in ${\bm E}^3$ are denoted with bold letters. Unless specified otherwise the associated reference frame is the inertial frame with respect to (w.r.t.) which the aircraft position is defined.
Inner and cross products in ${\bm E}^3$ are denoted by the symbols $\cdot$ and $\times$ respectively. Ordinary letters are used for real vectors of coordinates, and the $ith.$ component of a real vector $x$ is denoted as $x_i$.
The following notation is used.
\begin{itemize}
\item When $x \in \RR^n$ (resp. $\bm x \in {\bm E}^3$) $|x|$ (resp. $|\bm x|$) denotes the Euclidean norm of $x$ (resp. $\bm x$).
\item $\Piu$ denotes the operator of projection on the plane orthogonal to $\bm u$.
\item $\bar{\sat}^{\Delta}(y)$ ($\Delta>0$, $y \in \RR^n$) denotes a twice differentiable adaptation, with bounded derivatives, of the classical vector-valued saturation function $\sat^{\Delta}(y)=min(1,\frac{\Delta}{|y|})y$. More precisely $\bar{\sat}^{\Delta}(y)=\alpha^{\Delta}(|y|)y$ with $\alpha^{\Delta}:[0,+\infty)\mapsto (0,1]$ a decreasing twice differentiable function such that $\alpha^{\Delta}(0)=1$, $\frac{d}{dx}\alpha^{\Delta}(0)=\frac{d^2}{dx^2}\alpha^{\Delta}(0)=0$, $\alpha^{\Delta}(x)\leq \frac{\Delta}{x}$, $lim_{x\rightarrow +\infty}(\alpha^{\Delta}(x)x)=\Delta$. A typical example is $\alpha^{\Delta}(x)=\frac{\Delta}{x}\tanh(\frac{x}{\Delta})$. From these definitions $\bar{\sat}^{\Delta}(y)\approx y$ when $|y|$ is small and $|\bar{\sat}^{\Delta}(y)|\leq \Delta$, $\forall y$. The extension of this saturation function to an Euclidean vector $\bm y$ is $\bar{\sat}^{\Delta}(\bm y)=\alpha^{\Delta}(|\bm y|)\bm y$.
\item $G$ denotes the aircraft center of mass (CoM).
\item ${\mathcal I}=\{O;\bm \imath_0, \bm \jmath_0, \bm k_0 \}$ is an inertial frame with $\bm k_0$ pointing vertically and downward.
\item ${\mathcal B}=\{G;\bm \imath, \bm \jmath, \bm k\}$ is the chosen aircraft-fixed frame, with $\bm \imath$ and  $\bm \jmath$ parallel to the so-called zero-lift plane of the aircraft. The vector $\bm \imath$ (resp. $\bm \jmath$) is along the longitudinal (resp. lateral) motion direction of the aircraft (see Fig. \ref{frames}).
\item $\bm{\omega}$ is the angular velocity of ${\mathcal B}$ w.r.t. ${\mathcal I}$, i.e.
\begin{equation} \label{eq:newton0-2}
\frac{d}{dt} (\bm \imath,\bm \jmath,\bm{k}) = \bm{\omega} \times (\bm \imath,\bm \jmath,\bm{k})
\end{equation}
The vector of coordinates of $\bm{\omega}$ in the body-fixed frame ${\mathcal B}$ is denoted as $\omega$.
\item $m$ is the body mass.
\item $\bmp$ is the CoM position w.r.t. the inertial frame.
\item $\bm v$ is the CoM velocity w.r.t. the inertial frame, i.e.
\begin{equation} \label{velocity}
\dot{\bm p}=\bm v
\end{equation}
The vector of coordinates of $\bm v$ in the body-fixed frame ${\mathcal B}$ is denoted as $v$.
\item $\bm a$ ($=\dot{\bm v}$) is the CoM acceleration w.r.t. the inertial frame;
\item $\bm g =g_0 \, \bm k_0$ is the gravitational acceleration;
\item $\bm v_w$ is the ambient wind velocity w.r.t. ${\mathcal I}$, which we assume bounded and differentiable w.r.t. time;
\item $\bm v_a = \bm v-\bm v_w$ is the aircraft air-velocity. The vector of coordinates of $\bm v_a$ in the body fixed-frame ${\mathcal B}$ is denoted as $v_a$. 
\item The direction of $\bm v_a$ in the body frame is characterized by two angles $\alpha$ and $\beta$ such that
\begin{equation}
\label{def-ab}
\bm v_a = |v_a| (\cos \alpha( \cos \beta \, \bm \imath + \sin \beta \, \bm \jmath)+ \sin \alpha \, \bm k)
\end{equation}
$\alpha=\arcsin(v_{a,3}/|v_a| )$ and $\beta=\arctan(v_{a,2}/v_{a,1})$ denote the attack angle and sideslip angle respectively.
\end{itemize}
  \subsection{Aerodynamic forces}
\vspace{-0.2cm}
The resultant aerodynamic force ${\bm F}_a$ applied to a rigid body  moving with
air-velocity ${\bm v}_a$ is traditionally decomposed into the sum of a {\em drag} force ${\bm F}_D$ along the direction
of ${\bm v}_a$ and a {\em lift} force $\bm F_L$ perpendicular to this direction, i.e.
\begin{equation}
\label{fa}
\bm F_a= \bm F_D+ \bm F_L
\end{equation}
The intensities of drag and lift forces are essentially proportional to $|{v}_a|^2$ modulo variations characterized by
two dimensionless functions $C_D$ and $C_L$, which depend in the first place on the orientation of ${\bm v}_a$ w.r.t.
the body, but also on the Reynolds number $R_e$ and Mach number $M$.
These dimensionless functions are called the {\em aerodynamic characteristics} of the body, or {\em drag coefficient} and
{\em lift coefficient} respectively. More precisely
\begin{equation}
\label{fdl1}
{\bm F}_D={-}\eta_a |{v}_a| C_D \, {\bm v}_a~,~~{\bm F}_L=\eta_a |{v}_a| C_L \, {\bm v}_a^{\perp}
\end{equation}
with \begin{itemize}
\item $\bm v_a^{\perp}$ some vector perpendicular to $\bm v_a$ and such that
$|\bm v_a^{\perp}|=|v_a|$,
\item
$\eta_a := \frac{\rho \Sigma}{2}$ with $\rho$ the {\em free stream} air density, and $\Sigma$ an area germane to the body shape.
\end{itemize}

We neglect here the dependence of the aerodynamic characteristics on the Reynolds and Mach numbers. This is all the more justified for scale-model aircraft evolving at nearly constant altitudes and at speeds much lower than the speed of sound.
It is well known that the norm of aerodynamic forces is proportional to the squared norm of the air-velocity so that one can safely assume that there exists two positive numbers $c$ and $d$ such that
\begin{equation} \label{Fabound}
|\bm{F}_a|< c+d|v_a |^2
\end{equation}
The more specific model that we propose to use is:
\begin{equation} \label{Fa}
{\bm{F}_a}=-(c_0 v_{a,1} \bm \imath +\bar{c}_0 v_{a,3} \bm k)|v_a|+v_{a,2}\bm O(\bm v_a)
\end{equation}
with $c_0$ and $c_1$ denoting positive numbers, $\bar c_0=c_0+2c_1$, and $\bm O(\bm v_a)$ any Euclidean vector-valued function such that the ratio $\frac{|\bm O(\bm v_a)|}{|v_a|}$ is bounded. For instance, in \cite{KaiHamelSamsonCDC2017} we have used $\bm O(\bm v_a)=-c_0|v_a|\bm \jmath$ for a model of a disc-shaped aircraft. We voluntarily do not try to better specify the dependence of ${\bm{F}_a}$ upon the lateral airspeed $v_{a,2}$ because this dependence is both complex and less important than having a suitable model of the aerodynamic forces acting on the aircraft in the absence of lateral air-speed.
Note that relation \eqref{Fa} is compatible with the assumed aircraft symmetry about the plane $(G;\bm \imath,\bm k)$ and that, if the drag coefficient $c_0$ were equal to zero then, in the case of zero sideslip angle ($v_{a,2}=0$), the resultant aerodynamic force would be in the direction orthogonal to the zero-lift plane with an amplitude proportional to $\sin \alpha|v_a|^2$. This model must also be compatible with the general relations \eqref{fa} and \eqref{fdl1}, in particular when the lateral airspeed vanishes. In this latter case one easily verifies that the model \eqref{Fa} yields $\bm v_a^{\perp}=-\frac{|v_a|}{\cos \alpha}\bm k-\tan \alpha {\bm v_a}$, $C_D(\alpha)=(c_0+2 c_1 \sin^2\alpha)/\eta_a$, and $C_L(\alpha)=c_1 \sin2\alpha/\eta_a$.  For small attack angles $|\alpha|$ the drag coefficient $C_D$ is thus approximately equal to $\frac{c_0}{\eta_a}$ and the lift coefficient $C_L$ is approximately proportional to the attack angle with the coefficient of proportionality given by $\frac{2c_1}{\eta_a}$. This is coherent with experimental data performed on a variety of wing profiles and axisymmetric bodies \cite{pucci2015nonlinear}. Note also that, by contrast with other models, which are valid only locally for small attack angles, the proposed model respects the physical property of no lift when the air-velocity is perpendicular to the aircraft zero-lift plane, i.e. when $\alpha=\pi/2$. For control design purposes, using a model covering a large spectrum of operating conditions is of interest, especially for scale-model aircraft that are particularly sensitive to aerology conditions. Nevertheless, we must concede that this model is not enough elaborate to account for the abrupt and complex stall phenomena occurring beyond some attack angle (typically around $\alpha=\pi/10$). A way to tentatively overcome this shortcoming consists in modifying the coefficients $c_0$ and $c_1$ beyond the stall angle \cite{pucci2015nonlinear}.\\
\vspace{-0.2cm}
\subsection{Body dynamics}
\vspace{-0.2cm}
We assume that the control inputs consist of i) a thrust force $\bm T= T \bm \imath$ applied at $G$ along the zero-lift direction $\bm \imath$ (so that it does not produce a significant torque),
and ii) a torque vector $\bm{\Gamma}$. This torque can be produced in many ways. In the case of an airplane, it is essentially produced by the air deflected by the aircraft fixed and movable surfaces. In principle, movable control surfaces are dimensioned to produce torques that can dominate those produced by fixed surfaces. For this reason, these latter torques can, in the first approximation, be omitted in the modeling equations of the aircraft. Also, since moving surfaces are typically much smaller than fixed surfaces, the relative variation of aerodynamic forces about the neutral positions of these surfaces is limited and can also be neglected in the first approximation. Under these assumptions and approximations, the aircraft dynamic equations are given by \eqref{eq:newton0-2} complemented with the classical Newton-Euler equations:
\begin{equation} \label{eq:newton0-1}
m {\bm a} = m\bm{g} +\bm{F}_a +T\bm \imath
\end{equation}
\begin{equation} \label{eq:newton0-3}
J \dot{\omega} =-S(\omega)J\omega+\Gamma
\end{equation}
with $J$ the body inertia matrix, and $\Gamma$ the vector of coordinates of $\bm{\Gamma}$ in the body frame.\\
The {\em glide rate} $gr$ of an aircraft is commonly defined as the largest possible ratio between constant horizontal and vertical speeds of the aircraft in the vertical plane (i.e. the smallest rate of descent) when no thrust is applied. In the case of the dynamics model \eqref{Fa}, \eqref{eq:newton0-1} here considered one verifies that
\[
gr=\sup_{\alpha}\frac{{\bm F}_L(\alpha)}{{\bm F}_D(\alpha)}=\sup_{\alpha}\frac{C_L(\alpha)}{C_D(\alpha)}=\frac{1-c_0/\bar{c}_0}{2\sqrt{c_0/\bar{c}_0}}
\]
In the nominal case where $c_0/\bar{c}_0\ll 1$ this relation simplifies to
\begin{equation} \label{gr}
gr\approx 0.5\sqrt{\bar{c}_0/{c}_0}
\end{equation}
With the same approximation the corresponding gliding speed and sink rate are given by
\begin{equation} \label{vsink}
\begin{array}{l}
|v_{gr}|\approx \sqrt{m|\bm g|}/(c_0\bar{c}_0)^{0.25}\\
v_{sink}\approx v_{gr}/gr \approx 2\sqrt{mg_0}c_0^{0.25}/\bar{c}_0^{0.75}
\end{array}
\end{equation}
\\
In view of Eq.~\eqref{eq:newton0-3}, $\omega$ can be modified at will via the choice
of the control torque ${\Gamma}$ so that one can consider the angular velocity $\bm{\omega}$
as an intermediate control input. 
The corresponding physical assumption is that ``almost'' any desired angular velocity can
be obtained after a short transient time. This is a standard ``backstepping'' assumption.
Once it is made the body dynamic equations are reduced to \eqref{eq:newton0-2}, \eqref{velocity} and \eqref{eq:newton0-1}, with four input variables,
namely the thrust intensity $T$ and the three components of $\omega$. This assumption allows one to "eliminate" or, more precisely, postpone the complementary issue of producing a desired angular velocity via a torque $\Gamma$ that can be produced via various and specific physical means. In the case of a standard aircraft using the air deflected by movable control surfaces, a possible calculation of this torque, which does not take the contribution of fixed surfaces into account, is specified in Section \ref{control_surfaces}.
\vspace{-0.2cm}
\section{Control}
\label{control}
\vspace{-0.2cm}
\begin{figure}
 \includegraphics[width=.47\textwidth]{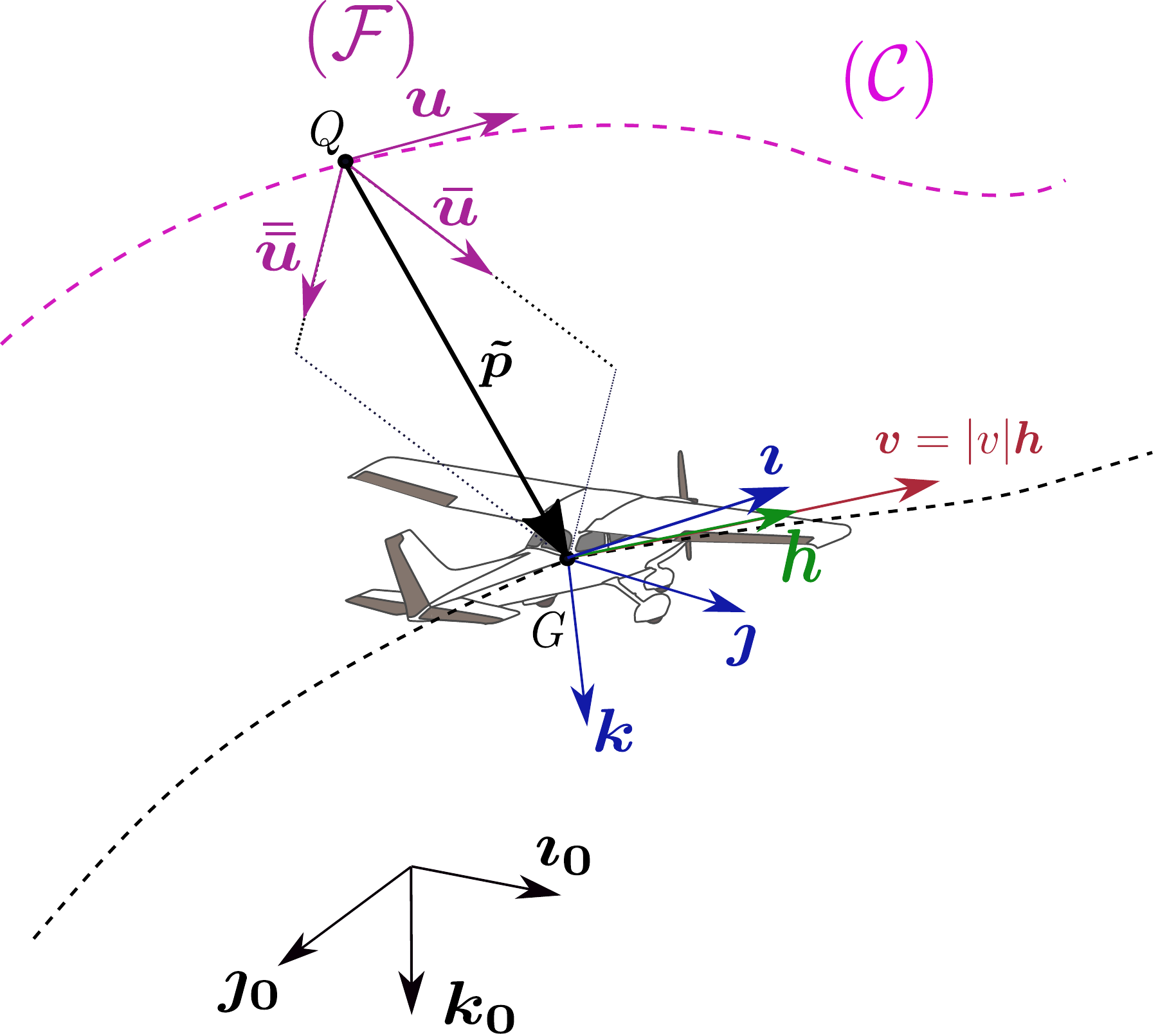}
 \caption{Desired path and position errors}
 \label{pathfoll}
\end{figure}
\subsection{Objectives and useful relations} \label{useful}
\vspace{-0.2cm}
Consider a three-times differentiable curve ${\mathcal{C}}$ in 3D-space parametrized by its curvilinear abscissa $s$ and, at the point $Q(s)$ on this path, an associated {\em parallel transport frame} ${\mathcal{F}}=\{Q;\bm u,\barbmu,\barbarbmu \}$ with
$\bmu$ the vector tangent to the path ${\mathcal{C}}$ at the point $Q$. An advantage of a parallel transport frame over the more conventional Frenet-Serret frame
is that it is (continuously) well-defined at points where the path curvature vanishes. It does not suffer from ambiguity and sudden orientation changes when the curves straightens out \cite{Bishop1975}\cite{Hanson95paralleltransport}. Its (relative) drawback is that it is not defined from the sole curve characteristics (curvature and torsion). More precisely, it is uniquely defined only once the vectors $\barbmu$ and $\barbarbmu$ are (arbitrarily) chosen at some point on the curve. The corresponding variational frame equations are
\begin{equation} \label{parallel-transport}
\frac{d}{ds}\barbmu=-\gamma_1\bmu~;~\frac{d}{ds}\barbarbmu=-\gamma_2\bmu~;~\frac{d}{ds}\bmu=\gamma_1\barbmu+\gamma_2\barbarbmu
\end{equation}
With this formalism any smooth curve is characterized by an initial point at $s=0$, the choice of a parallel transport frame at this point, and the functions $\gamma_1(s)$ and $\gamma_2(s)$. These functions are themselves related to the curve curvature $\kappa$ and torsion $\tau$ according to $\kappa=\sqrt{\gamma_1^2+\gamma_2^2}$ and $\tau=\frac{d}{ds}(\arctan(\frac{\gamma_2}{\gamma_1}))$. For instance, $(\gamma_1=0,\gamma_2=0)$ in the case of a straight line, $(\gamma_1=\frac{1}{r},\gamma_2=0)$ in the case of a circle with radius $r$, and $(\gamma_1(s)=\kappa \cos(\tau s +c),\gamma_2(s)=\kappa \sin(\tau s +c))$ --with $c$ denoting a constant depending on the choice of the initial frame-- in the case of a helix with constant curvature and torsion.

Given a curve ${\mathcal{C}}$, the path following control objective consists in i) stabilizing the aircraft speed at a desired value, and ii) having the aircraft converge to the curve and then follow it. Let $\bmq$ denote the position of the point $Q$ on the curve closest to the aircraft CoM. Depending on the curve, this point can be always unique (as in the case of a straight line) or only locally unique, depending on the position of the aircraft w.r.t. the curve. For instance, in the case of a circle, uniqueness is granted provided that $G$ does not belong to the circle axis passing through the origin and perpendicular to the circle's plane. Define the position error vector $\bmtildep =: \bmp - \bmq$, and let $v^* \in \RR^+-\{0\}$ denote the desired magnitude of the aircraft speed. A way to achieve the previously evoked control objectives consists in stabilizing $|v|-v^*$ and $\bmtildep$ at zero.

By definition of $Q$ (point on the curve closest to $G$) the vector $\bmtildep$ is perpendicular to the tangent to the curve at $Q$. It thus belongs to the plane $\{Q;\barbmu,\barbarbmu \}$. Let $y=[y_1,y_2]^\top \in \RR^2$ denote the vector of non-zero coordinates of $\bmtildep$ expressed in the basis of the parallel transport frame ${\mathcal{F}}=\{Q;\bm u,\barbmu,\barbarbmu \}$, i.e. $\bmtildep=y_1\barbmu +y_2 \barbarbmu$ with $y_1=: {\bmtildep} \cdot {\barbmu}$ and $y_2=: {\bmtildep} \cdot {\barbarbmu}$. The convergence of $\bmtildep$ to zero is equivalent to the convergence of $y$ to zero. With $\bmtildep$ taken as a Euclidean vector w.r.t. the reference frame ${\mathcal{F}}$ one can make this vector converge to zero by considering its variations w.r.t. to this frame. We will use the notation $\bmtildepF$ and $\dotbmtildepF$ when the vector $\bmtildep$ and its time-derivative are taken as Euclidean vectors in the reference frame ${\mathcal{F}}$. 
One has
\begin{equation} \label{dottildep}
\dotbmtildep=\frac{d}{dt}\vec{OG}-\frac{d}{dt}\vec{OQ}=\bm v -\dot{s}\bmu
\end{equation}
and
\begin{equation} \label{first_derivative}
\begin{array}{lll}
\dotbmtildepF&=&\dot{y}_1 \barbmu +\dot{y}_2 \barbarbmu\\
~&=&\Piu \bm v\\
~&=&\bm v -(\bm u \cdot \bm v)\bm u
\end{array}
\end{equation}
with
\[
\dot{y}_1=\bm v \cdot \barbmu~;~\dot{y}_2=\bm v \cdot \barbarbmu
\]
We have already established that
\[
\begin{array}{lll}
\dot{s}\bmu&=&\bm v-\dotbmtildep\\
~&=&\bm v-(\dotbmtildepF+y_1\dot{\barbmu}+y_2\dot{\barbarbmu})\\
~&=&\bm v-\dotbmtildepF+(\gamma_1y_1+\gamma_2y_2)\dot{s}\bmu
\end{array}
\]
The scalar product of both members of the previous equality with $\bmu$ yields
\begin{equation} \label{dots}
\dot{s}=\frac{(\bmu \cdot \bm v)}{1-\gamma_1y_1-\gamma_2y_2}
\end{equation}
For the sake of avoiding problems of little practical relevance we will assume from now on that the chosen path is such that $\gamma_1$ and $\gamma_2$ are uniformly bounded. The uniqueness of the projection of the aircraft CoM on the path is then granted provided that $1-\gamma_1y_1-\gamma_2y_2$ is larger than some positive constant. In the case of a straight line, for which $\gamma_1=\gamma_2=0$, this condition is thus satisfied independently of the aircraft position. If this condition is satisfied and $|v|$ is bounded, then  the time-derivative of $\bm u$ is also bounded.

\subsection{Control design}
\begin{figure}
 \includegraphics[width=.470\textwidth]{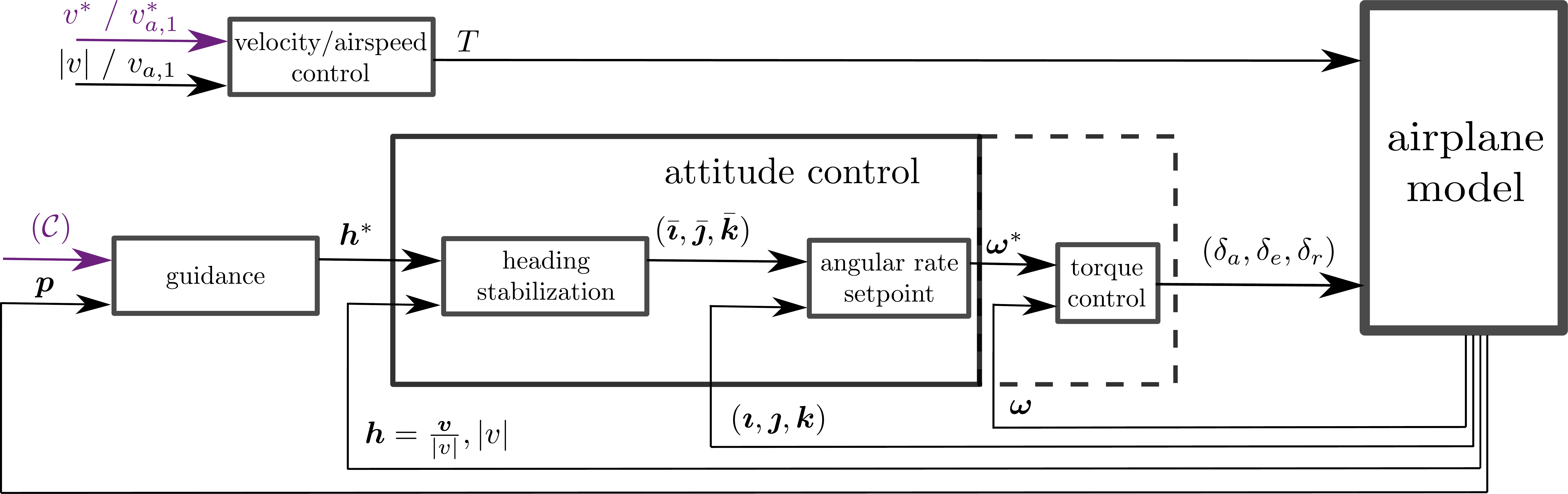}
 \caption{Control block diagram}
 \label{blockdiagram}
\end{figure}

\vspace{-0.2cm}
A central quantity involved in the control of any aircraft is the velocity vector $\bm v=|v|\bm h$ with $\bm h:=\frac{\bm v}{|v|}$ denoting the {\em heading vector} that specifies the direction of displacement of the aircraft, and $|v|$ the inertial speed. In the case of the path following problem here addressed the decomposition of $\bm v$ into the product of $|v|$ by $\bm h$ is all the more justified that convergence to the desired path can be performed at various speeds with the same heading policy.{\em Vice versa}, changing the heading policy does not imply modifying the aircraft speed. Decoupling the $|v|$ control problem from the control of $\bm h$ is also natural because these problems essentially involve separate physical inputs: monitoring of $|v|$ is done via thrust adaptation, i.e. via the control input $T$, whereas monitoring of $\bm h$ is performed via the control of the aircraft attitude, i.e. via the control input $\omega$.\\
The proposed control design involves the interconnection of three stages (see Fig. \ref{blockdiagram}), namely i) speed control via thrust adaptation, ii) desired heading vector determination (guidance), and iii) attitude control for the balanced flight stabilization of the desired heading vector. These three stages are detailed next.

\subsubsection{Speed control via thrust adaptation}
\vspace{-0.2cm}
Using the fact that $\bm v_a=v_{a,1}\bm \imath+v_{a,2}\bm \jmath+v_{a,3}\bm k$, the model of aerodynamic forces \eqref{Fa} may also be written as
\begin{equation} \label{Fa2}
{\bm{F}_a}=-\bar{c}_0 |v_a| \bm v_a+2c_1v_{a,1} |v_a|\bm \imath+mv_{a,2}\bm \bar{O}(\bm v_a)
\end{equation}
with $\bar{O}(\bm v_a)=\frac{1}{m}({O}(\bm v_a)+\bar{c}_0 |v_a|\bm \jmath)$. Combining this relation with \eqref{eq:newton0-1} yields
\begin{equation} \label{acceleration}
{\bm a} = \bm{\bar{g}} +\frac{\bar{T}}{m} \bm \imath+v_{a,2}\bm \bar{O}(\bm v_a)
\end{equation}
with
\begin{equation} \label{gbar}
\bm{\bar{g}}:=\bm{{g}}-\frac{\bar{c}_0}{m} |v_a| \bm v_a
\end{equation}
and
\begin{equation} \label{Tbar}
\bar{T}:=T+2c_1v_{a,1} |v_a|
\end{equation}
Let $v^*$ denote the desired value of $|v|$ and define $e_v:=|v|-{v^*}$. By using \eqref{acceleration} one finds that
\begin{equation} \label{inter1}
\frac{d}{dt}e_v=\bm{\bar{g}} \cdot \bm h + (\bm \imath \cdot \bm h) \frac{\bar{T}}{m}-\dot{v}^*+v_{a,2}\bm \bar{O}(\bm v_a) \cdot \bm h
\end{equation}
This relation in turn suggests to set
\begin{equation} \label{thrust1}
\bar{T}=m\big(-\bm{\bar{g}} \cdot \bm h+\dot{v}^*-k_{T,1}e_v-k_{T,2}\alpha_e I_{e_v}\big)/\bm \imath \cdot \bm h
\end{equation}
with $k_{T,1}$ and $k_{T,2}$ denoting positive gains, $I_{e_v}$ some bounded integral of the velocity error $e_v$ and the scalar function $\alpha_e$ defined by
\[
\alpha_e(e_v, I_{e_v}):=\alpha^{\Delta_{e_v}}(|I_{e_v}+e_v/k_{T,3}|)~~(\in (0,1])
\]
with $k_{T,3}$ denoting a positive number.
This Proportional-Integral (PI) feedback controller, complemented with a pre-compensation term, is well defined if $\bm \imath \cdot \bm v\neq0$ (a nominally satisfied condition) and yields the closed-loop equation
\begin{equation} \label{speed_error_eq}
\frac{d}{dt}e_v=-k_{T,1}e_v-k_{T,2} \alpha_e I_{e_v}+v_{a,2}\bm \bar{O}(\bm v_a) \cdot \bm h
\end{equation}
\begin{propo} \label{speed_control}
Assume that $\bm \imath \cdot \bm v$ stays larger than some positive number.
Choose control gains such that
\begin{itemize}
\item $k_{T,1}=k_{T,11}+k_{T,12}|e_v|^p$, with $k_{T,11}$ and $k_{T,12}$ denoting positive constants, and $p>1$;
\item the ratio $k_{T,1}/k_{T,2}$ is bounded from below and above.
\end{itemize} 
Define the bounded integral of $e_v$ as follows
\begin{equation} \label{Iev} 
\dot{I}_{e_v}=k_{T,2}k_{T,3}\big(-{I}_{e_v}+\bar{\sat}^{\Delta_{e_v}}({I}_{e_v}+e_v/k_{T,3})\big)
\end{equation}
with $k_{T,3}$ denoting a positive constant.
Then the application of the thrust control \eqref{thrust1} ensures the boundedness of $|e_v|$, and thus of $|v|$, uniformly with respect to the initial velocity. Furthermore, if the slideslip speed $v_{a,2}$ converges exponentially to zero, then $e_v$ also converges exponentially to zero. When $v_{a,2} \equiv 0$, the origin of the feedback controlled system \eqref{speed_error_eq} is exponentially stable.
\end{propo}
\vspace{-0.2cm}
How to control the aircraft attitude so as to make $v_{a,2}$ converge to zero, and thus achieve a so-called "balanced" flight with zero sideslip angle, is addressed later on via the control of the aircraft attitude. Note that, as pointed out in the proof of Proposition \ref{speed_control}, the gain $k_{T,12}$ is essentially "technically" useful to prove the boundedness of the aircraft velocity. Since it can be chosen arbitrarily small, it is in fact of limited practical value and may well be omitted in the control law, all the more so that the aircraft speed is physically bounded because of thrust limitations and energy dissipating drag forces.
Note also that the integral action is not only useful in practice to compensate for imprecisely known pre-compensation terms, but also to compensate for the imperfect knowledge of the ratio between the calculated desired thrust and the physically applied one.\\
\vspace{-0.2cm}
\subsubsection{Desired heading vector determination (guidance)}
\vspace{-0.2cm}
By viewing the aircraft as a point moving in 3D-space with a given speed $|v|$ the problem is to determine a desired heading direction $\bm h^*$ that yields the convergence of this point to the desired path and ensures that the point moves thereafter along the path with the desired direction given by $\pm \bm u$, i.e. such that $\bm h^*$ converges to $sign_{v_u}\bm u$ with $sign_{v_u}$ chosen in advance and equal to either $1$ or $-1$. There are obviously a multitude of solutions to this problem, a certain number of which have been reported in the literature \cite{BreivikFossen2005,PSKB2017,OAE2016,cichellaetal2011}. The solution that we propose here consists in setting
\begin{equation} \label{hstar}
\bm h^*=\sin(\theta_h)\bm l+\cos(\theta_h)sign_{v_u}\bm u
\end{equation}
with $\bm l$ denoting some unit vector orthogonal to $\bm u$ and $\theta_h$ an angle depending on the position error $\bmtildep$ and converging to zero when $|\bmtildep|$ tends to zero. For instance, a simple possible choice for $\bm l$ and $\theta_h$ is
\begin{subequations} \label{ybarl}
\begin{align}
&\bar{y}:=k_1D\bar{\sat}^{\Delta_{h}}(y)/|v| \label{ybarl:first}\\
&\theta_h:\arctan(|\bar{y}|/\sqrt{1-|\bar{y}|^2}) \label{ybarl:second}\\
&\bm l:=-\frac{\bar{y}_1\barbmu+\bar{y}_2\barbarbmu}{|\bar{y}|} \label{ybarl:third}
\end{align}
\end{subequations}
with $D=diag\{d_1,d_2\}$ a diagonal matrix with $d_i\in (0,1]$ ($i=1,2$), $\Delta_h:=\frac{\mu|v|}{k_1\max(d_1,d_2)}$, $\mu\in(0,1)$, and $k_1$ a positive gain. Note that $|\bar{y}|<1$ so that $\theta_h$ is well defined and belongs to $[0,\pi/2)$. Note also that $\sin(\theta_h)=|\bar{y}|\leq \mu$ so that the term $\sin(\theta_h)\bm l$ entering the expression of $\bm h^*$ is always well defined. Moreover, if $d_1=d_2$ then $\sin(\theta_h)$ tends to $\mu$ when $|y|$ tends to infinity. Therefore, in this case $\arcsin(\mu)$ characterizes the angle of incidence of the desired heading direction w.r.t. the tangent to the desired path at the point $Q$, when the aircraft is far from the path. The usefulness of choosing $d_1\neq d_2$ is related to the possibility of imposing different upper bounds upon $|\dot{y}_1|$ and $|\dot{y}_2|$, a feature which may be useful to separate the rates of convergence along the directions $\barbmu$ and $\barbarbmu$, and limit the rate of descent or of climb of the aircraft when, for instance, $\barbarbmu$ is a vertical vector as in the case of a linear or circular horizontal path.\\
The following proposition summarizes the stability and convergence properties associated with this desired heading direction.
\begin{propo} \label{stabilisation}
Assume that $1-\gamma_1y_1-\gamma_2y_2$ and $|v|$ stay larger than some positive number so that the position error $\tilde{\bm p}$ and the aircraft heading vector $\bm h$ are well defined at any time-instant. Assume that $|v|$ is bounded and that $\bm h=\bm h^*+\bm o$, with $\bm o$ denoting a "residual" vector such that the integral $\int_0^t |\bm o(s)|ds$ is bounded.\\
Then $|y|$ converges to zero, and $\bm h$ tends to $sign_{v_u}\bm u$. The rate of convergence is ultimately exponential if the rate of convergence of $|\bm o|$ is itself ultimately exponential. Moreover $|\dot{y}|$ is ultimately upper bounded by $\mu|v|$ and $|\dot{y}_i|$ is ultimately upper bounded by $\frac{d_i}{\max(d_1,d_2)}\mu |v|$ ($i=1,2$). In the case where $\bm o \equiv \bm 0$ the equilibrium $\tilde{\bm p}=\bm 0$ is locally exponentially stable.\\~\\
 \end{propo}
\vspace{-0.5cm}
{\bf Remark:} Any other guidance law $\bm h^*$ yielding the same properties when $\bm h=\bm h^*$, i.e. local exponential stability of $\tilde{\bm p}=\bm 0$ and convergence of $\bm h^*$ to $sign_{v_u}\bm u$, can be used in combination with the attitude controller derived next. Alternative guidance laws are for instance proposed in \cite{park2004new,BreivikFossen2005,cichellaetal2011}.
\subsubsection{Attitude control}
\vspace{-0.2cm}
Attitude control is in charge of making the aircraft heading direction $\bm h$ converge to the desired one $\bm h^*$ and of ensuring a balanced flight, i.e. of zeroing the side-slip angle by zeroing the lateral velocity component $v_{a,2}=\bm v_a.\bm \jmath$. We show next that these two objectives can be achieved via the determination of a desired mobile frame ${\bar{\mathcal B}}:=\{G;\bm {\bar{\imath}}, \bm {\bar{\jmath}}, \bm {\bar{k}}\}$ and the convergence of the aircraft frame ${\mathcal B}=\{G;\bm {{\imath}}, \bm {{\jmath}}, \bm {{k}}\}$ to this desired frame. 
Let $\bm \omega_h$ denote the angular velocity of $\bm h$, i.e. 
\begin{equation} \label{omegh}
\dot{\bm h}=\bm \omega_h \times {\bm h}
\end{equation}
and $\bm \omega_h=\bm h \times \dot{\bm h}$. Let $\bar{\bm \omega}_h$ denote a "desired" angular velocity for the heading vector $\bm h$ that ultimately exponentially stabilizes $\bm h=\bm h^*$ when $\bm \omega_h=\bar{\bm \omega}_h$. Take, for instance
\begin{equation} \label{omegah}
\bar{\bm \omega}_h:=\bm \omega_{h^*}+k_{h}\tilde{\bm h}
\end{equation}
with $\bm \omega_{h^*}:=\bm h^* \times \dot{\bm h}^*$ denoting the angular velocity of $\bm h^*$, $\tilde{\bm h}:=\bm h \times \bm h^*$, and $k_{h}$ a positive gain (not necessarily constant) whose value determines the rate of convergence of $\bm h$ to $\bm h^*$. The almost global asymptotic (locally exponential) stability of $\bm h=\bm h^*$ when $\bm \omega_h=\bar{\bm \omega}_h$ then results from that $\frac{d}{dt}(1-\bm h \cdot \bm h^*)=-k_h|\tilde{h}|^2~\leq 0$. The domain of stability is not global because ${\bm h}=-\bm h^*$ is also an equilibrium. The instability of this equilibrium, and convergence to the other equilibrium when ${\bm h}$ is initially different from $-\bm h^*$, come from that the non-increasing function $1-\bm h \cdot \bm h^*$ has its maximal value (equal to $2$) at this equilibrium.\\
A more complete solution involves a complementary integral action in charge of compensating for stationary effects of errors in the modeling of the aircraft dynamics that could prevent the convergence of $\bm h$ to $\bm h^*$.
Without a complementary integral action, in the absence of wind, and along simple desired paths corresponding to classical {\em trim} trajectories (straight lines, horizontal circles,...), the error vector $\tilde{\bm h}$ would typically converge to a constant vector w.r.t a frame centered on the aircraft CoM and rotating with the angular velocity $\bm \omega_{h^*}$ of $\bm h*$. This suggests to use a bounded integral term calculated according to
\begin{equation} \label{saturated_integral}
\dot{\bm z}=\bm \omega_{h^*} \times \bm z
+k_z\big(-\bm z+\bar{\sat}^{\Delta_z}(\bm z+\tilde{\bm h} /k_z)\big)~;~\bm z(0)=\bm 0
\end{equation}
with $\Delta_z>0$ the chosen upper bound for $|z(t)|$, and $k_z$ denoting a positive number. The expression \eqref{omegah} of $\bar{\bm \omega}_h$ is then modified to
\begin{equation} \label{omegah2}
\bar{\bm \omega}_h:=\bm \omega_{h^*}+k_{h,1}\tilde{\bm h}+k_{h,2}\alpha_h \bm z
\end{equation}
with $k_{h,1}$ denoting a positive gain and the scalar function $\alpha_h$ defined by
\[
\alpha_h(\tilde{\bm h},\bm z):=\alpha^{\Delta_z}(|\bm z+\tilde{\bm h} /k_z|)~~(\in (0,1])
\]
From now on, the arguments of this function are omitted for the sake of legibility.
\begin{propo} \label{hstabilisation}
Assume that $1-\gamma_1y_1-\gamma_2y_2$ and $|v|$ stay larger than some positive number so that the position error $\tilde{\bm p}$, the desired heading vector $\bm h^*$, and the aircraft heading vector $\bm h$ are well defined at any time-instant.
Assume also that $\bm \omega_h=\Pi_{\bm h}\bar{\bm \omega}_h+\bm o$, with $\bar{\bm \omega}_h$ given by \eqref{saturated_integral}-\eqref{omegah2} and $\bm o$ a "residual" vector such that the integral $\int_0^t |\bm o(s)|ds$ is bounded. We distinguish two cases:\\~\\
\noindent {\bf case 1}:~$\forall t:~\bm o(t)=\bm 0$.\\
\begin{itemize}
\item In this case, the system \eqref{omegh}, \eqref{saturated_integral} has two equilibria, namely $(\bm h,\bm z)=(\bm h^*,\bm 0)$ and $(\bm h,\bm z)=(-\bm h^*,\bm 0)$. The first of these equilibria is locally exponentially stable, whereas the second one is unstable; 
\item $(\bm h, \bm z)$ converges to the first (desired) equilibrium provided that $\bm h(0) \neq -\bm h^*(0)$.
\end{itemize}
\noindent {\bf case 2}:~$\exists t:~\bm o(t)\neq\bm 0$. 
\begin{itemize} 
\item If $(\bm h,\bm z)$ does not converge to the unstable asymptotic equilibrium $(-\bm h^*,\bm 0)$ then it converges to the desired asymptotic equilibrium $(\bm h^*,\bm 0)$.
\item If $|\bm o|$ converges ultimately exponentially to zero, then the convergence of $(\bm h,\bm z)$ to $(\bm h^*,\bm 0)$ is also ultimately exponential.
\end{itemize}
\end{propo}
Let us now define the desired mobile frame ${\bar{\mathcal B}}$. By differentiating both members of the equality
$\bm v=|v|\bm h$ w.r.t. time, one deduces that
\[
\begin{array}{lll}
\bm a &=& \dot{|v|}\bm h+|v|\dot{\bm h}\\
~&=&\dot{|v|}\bm h+|v|({\bm \omega}_h \times \bm h)
\end{array}
\]
This relation suggests to define a "desired" acceleration as follows
\begin{equation} \label{astar}
\bm a^*:=\dot{v}^*\bm h+|v|(\bar{\bm \omega}_h \times \bm h)
\end{equation}
with $\bar{\bm \omega}_h$ given by \eqref{saturated_integral}, \eqref{omegah2}.
From relation \eqref{acceleration} we note that
\begin{equation} \label{bmi}
\bm \imath=\frac{{\bm a} -\bm{\bar{g}} -v_{a,2}\bm \bar{O}(\bm v_a)}{|{\bm a} -\bm{\bar{g}} -v_{a,2}\bm \bar{O}(\bm v_a)|}
\end{equation} 
The desired acceleration is in turn used to define $\bm {\bar{\imath}}$ as follows (compare with \eqref{bmi})
\begin{equation} \label{ibar}
\bm {\bar{\imath}}:=\frac{{\bm a^*} -\bm{\bar{g}}}{|{\bm a^*} -\bm{\bar{g}}|}
\end{equation}
For the vector $\bm {\bar{\jmath}}$ we set
\begin{equation} \label{jbar}
\bm {\bar{\jmath}}:=\frac{\bm v_a \times \bm {\bar{\imath}}}{|\bm v_a \times \bm {\bar{\imath}}|}=\frac{\bm v_a \times ({\bm a^*} -\bm{{g}})}{|\bm v_a \times ({\bm a^*} -\bm{{g}})|}
\end{equation}
and the third vector $\bm {\bar{k}}$ is just calculated as the cross product of $\bm {\bar{\imath}}$ and $\bm {\bar{\jmath}}$, i.e.
\begin{equation} \label{kbar}
\bm {\bar{k}}:=\bm {\bar{\imath}} \times \bm {\bar{\jmath}}
\end{equation}
An important property is that, like $\bm {\tilde{p}}$, $\bm v$, $\bm v_a$, $\bm g$, $\bm \omega_h^*$ and $\bm \omega_h$, the unit vectors $(\bm {\bar{\imath}},\bm {\bar{\jmath}},\bm {\bar{k}})$ so defined do not depend on the aircraft attitude. Therefore, their time-derivatives do not depend on the aircraft angular velocity $\bm \omega$.
Let $\bm \omega_{\bar{\bm \imath}}:=\bar{\bm \imath} \times \dot{\bar{\bm \imath}}$ and $\bm \omega_{\bar{\bm \jmath}}:=\bar{\bm \jmath} \times \dot{\bar{\bm \jmath}}$ denote the angular velocities of $\bm {\bar{\imath}}$ and $\bm {\bar{\jmath}}$ respectively. The angular velocity of the frame $\bar{\mathcal B}$ is then given by $ \bar{\bm \omega}=\bm \omega_{\bar{\bm \imath}}+(\bar{\bm \imath}.\bm \omega_{\bar{\bm \jmath}})\bar{\bm \imath}=\bm \omega_{\bar{\bm \jmath}}+(\bar{\bm \jmath}.\bm \omega_{\bar{\bm \imath}})\bar{\bm \jmath}$, and this vector does not depend on $\bm \omega$ either. The problem of stabilizing ${\bar{\mathcal B}}={\mathcal B}$ is thus well-posed.
\begin{propo} \label{framestabilisation}
Assume that $1-\gamma_1y_1-\gamma_2y_2$ and $|v|$ stay larger than some positive number so that the position error $\tilde{\bm p}$, the aircraft heading vector $\bm h$, and the desired heading vector $\bm h^*$ are well defined at any time-instant.
Further assume that $|{\bm a^*} -\bm{\bar{g}}|$, $|\bm {\bar{\imath}} \times \bm v_a|$, remain larger than some positive number so that the frame ${\bar{\mathcal B}}$ and its angular velocity $ \bar{\bm \omega}$ are also well-defined. Then an angular velocity control that almost globally asymptotically (locally exponentially) stabilizes ${\mathcal B}={\bar{\mathcal B}}$ is
\begin{equation} \label{omega}
\bm \omega=\bar{\bm \omega}+k_{\omega}(t)\big((\bm \imath \times \bar{\bm \imath})+(\bm \jmath \times \bar{\bm \jmath})+(\bm k \times \bar{\bm k})\big)
\end{equation}
with $k_{\omega}(t)>\epsilon>0$.
\end{propo}
\vspace{-0.2cm}
Prior to stating an overall stability result, the following proposition summarizes convergence properties that can be established from the partial results obtained so far.
\begin{theo}[Convergence] \label{hstabilisation2}
Consider an aircraft whose motion equations satisfy the kinematic equations \eqref{eq:newton0-2}-\eqref{velocity} and the Newton dynamic equation \eqref{eq:newton0-1}, complemented with the model \eqref{Fa} of aerodynamic forces. Given a desired heading vector $\bm h^*$, apply to this system
the attitude angular velocity control \eqref{omega}, combined with the thrust control defined by \eqref{Tbar}\eqref{thrust1}\eqref{Iev}.
Assume that during the flight $1-\gamma_1y_1-\gamma_2y_2$, $|v|$, $|\bm a - \bm{\bar{g}}|$, $|{\bm a^*} -\bm{\bar{g}}|$, $|\bm h \cdot (\bm a - \bm{\bar{g}})|$, $|\bm {\bar{\imath}} \times \bm v_a|$, and $|\bm \imath \cdot \bm v|$ stay larger than some positive number. Then $|v|$ converges to $v^*$. Provided that $\tilde{\theta}(0)$, i.e. the initial angle between the aircraft frame and $\bar{\mathcal B}$, is different from $\pi$, the aircraft frame converges to $\bar{\mathcal B}$ and the sideslip angle converges to zero.
Furthermore, if $(\bm h,z)$ does not converge to the unstable point $(-\bm h^*,0)$, then $(\bm h,z)$ converges to $(\bm h^*,0)$.
In this latter case, if $\bm h^*$ is given by \eqref{hstar}-\eqref{ybarl}, then the path following error $\tilde{\bm p}$ converges to zero and $\bm h$ converges to the desired direction, i.e. $sign_{v_u}\bm u$. Rates of convergence to the desired equilibria are ultimately exponential.
\end{theo}
\vspace{-0.2cm}
The conditions pointed out in Theorem \ref{hstabilisation2}, under which convergence to the desired path is granted, may seem restrictive at first glance; but they are in fact inherent to the control problem at hand. They are also related to the existence of particular trajectories along which the linearized equations of the system are not controllable. Although they are not satisfied in only very specific situations, they nonetheless rule out the possibility of global convergence results. However, it remains possible to state local asymptotic stability results when these conditions are satisfied on the desired path. For instance, in the case of zero wind velocity, and when $|v|=v^*$ is constant, one verifies that these conditions are satisfied on the desired path if $|\bm u \times \bm \imath|$ and $|\bm u.\bm \imath|$ are positive (and larger than a small number) on the path. Since, for a balanced flight, $\bm \imath=\frac{\bm a -\bm{\bar{g}}}{|\bm a -\bm{\bar{g}}|}$ with $\bm a={v^*}^2(\gamma_1\barbmu +\gamma_2\barbarbmu)$ and $\bm{\bar{g}}=\bm g-\frac{\bar{c}_0}{m}{v^*}^2 \bm u$, these conditions are themselves satisfied if
\begin{equation} \label{A1}
\mbox{\bf A1:}~|\bm g \times \bm u - {v^*}^2(\gamma_1\barbarbmu -\gamma_2\barbmu)|>\epsilon_1>0
\end{equation}
and
\begin{equation} \label{A2}
\mbox{\bf A2:}~|\frac{\bar{c}_0}{m}{v^*}^2-\bm g \cdot \bm u|>\epsilon_2>0
\end{equation}
We can then state the following local exponential stability result
\begin{theo}[Local exponential stability] \label{local_stability}
Given the model \eqref{Fa} of aerodynamic forces, consider the control system composed of the kinematic equations \eqref{eq:newton0-2}-\eqref{velocity} and the Newton dynamic equation \eqref{eq:newton0-1}, augmented with the integrators \eqref{Iev} and \eqref{saturated_integral}.
In the case of zero wind velocity and $v^*$ constant ($\neq 0$), if the assumptions A1-A2 are satisfied and $\bm h^*$ is a desired heading vector defined by \eqref{hstar}-\eqref{ybarl},
then the control inputs $(T,\bm \omega)$ defined by \eqref{Tbar},\eqref{thrust1},\eqref{Iev} and \eqref{omega} locally exponentially stabilize the equilibrium $(\bm p,\bm v,{\mathcal B},\bm z,I_{e_v})=(\bm q,sign_{v_u}v^*\bm u,\bar{\mathcal B},\bm 0,0)$.
\end{theo}
\vspace{-0.3cm}
{\bf Remark}: Despite the no-wind and constant desired speed assumptions, Assumptions A1-A2 are less conservative than the convergence conditions of Theorem \ref{hstabilisation2} from which they derive, because they bear only upon the desired aircraft trajectory and speed. On the other hand, Theorem \ref{hstabilisation2} shows that convergence is possible even when the aircraft starts far away from the desired equilibrium. Non-satisfaction of the convergence conditions over a long period of time is a remote possibility in practice, even in the absence of a good control. Nevertheless, this possibility cannot be discarded {\em a priori}. It thus matters to take practical precautions and implement control expressions which, besides from being efficient during the time-periods when these conditions are met, yield control inputs that are well defined and bounded in all circumstances. In particular, of course, no division by zero should be allowed. 
\vspace{-0.2cm}
\subsection{Application to particular curves} \label{particular_curves}
\vspace{-0.2cm}
\subsubsection{Straight line} \label{straight_line}
\vspace{-0.2cm}
$\mathcal {C}$ is a straight line passing through the point $\bm p_c$ and with constant unit direction vector $\bm u$. Then $\gamma_1=\gamma_2=0$. Assumption A1 is verified provided that the path is not vertical, i.e. not parallel to the gravitational acceleration. As for Assumption A2, it gives a condition relating the desired speed to the path slope. More precisely it is verified when ${v^*}^2\neq\frac{mg_0}{\bar{c}_0}\sin(\nu)$, with $\nu$ denoting the so-called path angle, i.e. the angle between the path and the horizontal plane. Moreover, the point $Q$ is always unique and its position can be directly calculated from the aircraft position $\bm p$ and the curve characteristics $(\bm p_c,\bmu)$. More precisely, $\bm q=\bm p_c+\big(\bmu.(\bm p-\bm p_c)\big)\bmu$, $\bmtildep=\bmu \times \big((\bm p-\bm p_c)\times \bmu\big)$ and any pair $(\barbmu,\barbarbmu)$ of constant orthonormal vectors perpendicular to $\bm u$ can be used for the control calculations.
\vspace{-0.2cm}
\subsubsection{Circle} \label{circle}
\vspace{-0.2cm}
$\mathcal {C}$ is a circle centered at $\bm p_c$ with radius $r$ and constant unit vector $\barbarbmu$ orthogonal to the circle's plane. Note that this plane does not have to be horizontal. Then $\gamma_1=\frac{1}{r}$, $\gamma_2=0$. As for the straight line case, the point $Q$ on the curve and the unit vectors $(\barbmu,\bmu)$ associated with the parallel transport frame (that coincides in this particular case with the Frenet-Serret frame) can be directly calculated from the aircraft position $\bm p$ and the curve characteristics $(\bm p_c,r,\barbarbmu)$. More precisely, $\barbmu=\frac{\big((\bm p-\bm p_c) \times \barbarbmu\big)\times \barbarbmu}{|\big((\bm p-\bm p_c)\times \barbarbmu\big)\times \barbarbmu|}$, $\bmu=\barbmu \times \barbarbmu$, $\bm q=\bm p_c -r \barbmu$, and $\bmtildep=\bm p -\bm q$. 
\newline The condition of positivity of $(1-\gamma_1y_1-\gamma_2y_2)$ ensuring the good conditioning of the projection of the aircraft CoM on the circle may also be written as $(\bm p -\bm p_c).\barbmu<0$. It is not satisfied only when the aircraft is located on the circle's axis, which corresponds to the case where the aircraft is equidistant to all points on the circle (loss of uniqueness of the closest point). Assumption A1 is always verified, except in the particular case when the circle is vertical and $g_0=\frac{{v^*}^2}{r}$. As for Assumption A2, it is verified when ${v^*}^2>\frac{mg_0}{\bar{c}_0}\sin(\nu)$, with $\nu$ denoting the angle between the circle's plane and the horizontal plane.
\vspace{-0.2cm}
\section{Complementary issues} \label{complementary}
\vspace{-0.2cm}
\subsection{Extension to a curve defined on a translating frame}
\vspace{-0.2cm}
An extension of the path following problem addressed previously consists in considering a curve that is fixed w.r.t. a translating frame whose origin $\bm p_c$ moves with velocity $\bm v_c$, i.e. $\frac{d}{dt}\bm p_c=\bm v_c$, and acceleration $\bm a_c$, i.e. $\frac{d}{dt}\bm v_c=\bm a_c$. For the sake of simplification we assume here that this frame does not rotate, but the extension to a rotating frame is also possible. A practical application would, for instance, consist in having an aircraft fly in circles over a moving ground target. This problem is also referred to as Moving Path Following (MPF) in \cite{OAE2016}\cite{Oliveira2017}. We leave the interested reader to verify that this extension just involves the modification of equations
\eqref{dottildep}, \eqref{first_derivative}, \eqref{dots}
according to
\begin{equation} \label{dottildep2}
\dotbmtildep=\frac{d}{dt}\vec{OG}-\frac{d}{dt}\vec{OC}-\frac{d}{dt}\vec{CQ}=(\bm v-\bm v_c) -\dot{s}\bmu
\end{equation}
\begin{equation} \label{first_derivative2}
\begin{array}{lll}
\dotbmtildepF&=&((\bm v-\bm v_c) \cdot \barbmu)\barbmu +((\bm v-\bm v_c)\cdot\barbarbmu)\barbarbmu\\
~&=&\Piu (\bm v-\bm v_c)
\end{array}
\end{equation}
\begin{equation} \label{dots2}
\dot{s}=\frac{((\bm v-\bm v_c) \cdot \bmu)}{1-\gamma_1y_1-\gamma_2y_2}
\end{equation}
The desired heading direction $\bm h^*_c$ ($=\frac{\bm v}{|v|}$) must satisfy the equation
\[
\frac{\bm v-\bm v_c}{|v-v_c|}=\bm h^*~(=\frac{\bm h^*_c-\frac{\bm v_c}{|v|}}{|\bm h^*_c-\frac{\bm v_c}{|v|}|})
\]
with $\bm h^*$ given by \eqref{hstar}. One verifies that the solution to this equation is
\begin{equation} \label{generalhstar}
\bm h^*_c=\Pi_{\bm h^*}\bm v_c/|v|+\sqrt{1-|\Pi_{\bm h^*}\bm v_c|^2/|v|^2}\bm h^*
\end{equation}
with $\Pi_{\bm h^*}$ denoting the operator of projection on the plane orthogonal to $\bm h^*$. For this desired direction to be well defined it suffices that $|v|>|v_c|$. As expected, $\bm h^*_c=\bm h^*$ when $\bm v_c=0$.
\vspace{-0.2cm}
\subsection{Airspeed control}
\vspace{-0.2cm}
Instead of $|v|$, one may wish to monitor the air-velocity in the direction of $\bm \imath$, i.e. the component $v_{a,1}=\bm v_a \cdot \bm \imath$, which can be measured, for instance, with a Pitot tube. Define now the speed error as $e_v:=v_{a,1}-v^*$. Using \eqref{acceleration} the time-derivative of $e_v$ satisfies the relation
\[
\begin{array}{lll}
\frac{d}{dt}e_v&=&\frac{d}{dt}((\bm v-\bm v_w) \cdot \bm \imath)-\dot{v}^*\\
~&=&(\bm a-\dot{\bm v}_w) \cdot \bm \imath+\bm \omega  \cdot (\bm \imath \times \bm v_a)-\dot{v}^*\\
~&=&(\bm g-\dot{\bm v}_w) \cdot \bm \imath+\bm \omega \cdot (\bm \imath \times \bm v_a)-\frac{c_0}{m}|v_a|v_{a,1}\\
~&~&-\dot{v}^*+\frac{T}{m}+v_{a,2}\bar{O}(\bm v_a) \cdot \bm \imath
\end{array}
\]
Assuming that $v_{a,2}$ converges exponentially to zero (balanced flight), exponential convergence of $v_{a,1}-v^*$ to zero is then obtained by setting
\begin{equation} \label{thrust3}
T=T^*-m(k_{T,1}e_v+k_{T,2}\alpha_eI_{e_v})
\end{equation}
with $T^*:=m(\dot{v}^*-(\bm g-\dot{\bm v}_w) \cdot \bm \imath-\bm \omega  \cdot (\bm \imath \times \bm v_a))+{c_0}|v_a|v_{a,1}$, $k_{T,1}$ and $k_{T,2}$ denoting positive gains, and $I_{e_v}$ some bounded integral of $e_v$ calculated, for instance, according to \eqref{Iev}.
 \vspace{-0.2cm}
\subsection{Adaptation to two-axis autopilots} \label{twoaxis}
\vspace{-0.2cm}
The three-axis (pitch-roll-yaw) control system proposed so far does not specifically rely on the existence of a stabilizing tail equipped with pitch and yaw control surfaces. All that matters is the possibility of modifying the aircraft attitude at will, like multirotor drones that use differential blade rotation to this purpose. By contrast, two-axis (either pitch-roll or pitch-yaw) control systems rely on the existence of a stabilizing tail that compensates, via passive control and torque coupling, for the lack of an active control torque about one of the aircraft rotation axes. Typically tail vertical and horizontal surfaces serve to stabilize the aircraft heading along the air-velocity direction by creating passive torques that are opposed to angular variations w.r.t. the air-velocity direction.
In particular, provided that an adequate roll angle is created, the tail vertical surface is very efficient at maintaining the side-slip angle small. This explains why active yaw control (often termed as yaw damping in relation to the comfort it brings to passengers by allowing for precise side-slip angle zeroing) via the use of a rotating rudder surface is of secondary importance for most common airplanes. This also explains the common use of two-axis pitch-roll autopilots that control airplanes automatically. With respect to the three-axis desired angular velocity $\omega^{\star}$ given by \eqref{omega} it thus essentially suffices to create, via elevator and ailerons actions, pitch and roll torques that asymptotically stabilize $\omega_1-\omega^{\star}_1$ and $\omega_2-\omega^{\star}_2$ at zero. This can be achieved without creating a yaw torque with the tail rudder.
\vspace{-0.2cm}
\subsection{Thrust bounds and attack angle monitoring}
\vspace{-0.2cm}
We have so far assumed that the aircraft could produce the desired thrust $T$ calculated according to \eqref{Tbar}-\eqref{thrust1}. In practice this desired value of $T$ may leave the physical thrust interval $[T_{min},T_{max}]$. When this happens at least one of the control objectives --i.e. convergence of the aircraft heading direction $\bm h$ to the desired one $\bm h^*$ (or $\bm h^*_c$), or stabilization of $|v|$ at the desired speed $v^{\star}$-- cannot (momentarily) be achieved with the available thrust. For instance, descending from a high altitude to a horizontal circular path with the convergence dynamics specified by $\bm h^*$ may require a negative thrust (to slow down the aircraft) that a common aircraft, for which $T_{min} \approx 0$, cannot produce. Similarly, ascending with these convergence dynamics and velocity may require a thrust that exceeds $T_{max}$.\\
To avoid this situation a possibility consists in reducing the rate of convergence to the desired path in order to have the control action focused on the stabilization of the aircraft speed about the desired speed $v^*$. This can be done, for instance, by choosing the parameter $\mu$ in the expression of the saturation $\Delta_h:=\frac{\mu|v|}{k_1max(d_1,d_2)}$ small enough. Provided that $v^*$ can be maintained on the desired path with the available thrust (meaning for instance, in the case of a straight path and assuming that $(-sign_{v_u}m\bm g.\bm u+c_0{v^*}^2)\in[T_{min},T_{max}]$), this strategy can ensure that the calculated thrust $T$ will ultimately remain in the interval $[T_{min},T_{max}]$ whatever the initial position and orientation of the aircraft. In the case of a horizontal path one can alternatively set the parameter $d_2$ small enough so as to impose a (small) rate of climb or descent, during the transient phase of convergence to the path, that is compatible with the thrust limitations.\\
The aforementioned possibility aims at keeping the calculated thrust within the range $[T_{min},T_{max}]$ so that both control objectives can simultaneously be achieved. However, one may also temporarily accept an increase of the aircraft speed beyond the desired speed $v^*$ during a descending transient phase when a negative thrust (needed to slow down the aircraft) is calculated and cannot be produced (i.e. when $T_{min}=0$). In this case only the objective of convergence of the heading direction $\bm h$ to the desired one $\bm h^*$ is maintained. This implicitly means that the current aircraft speed obtained with zero thrust (the saturated value closest to the calculated negative thrust) coincides with the desired speed, i.e. $v^*(t)=|v(t)|$. This in turn yields to setting $\dot{v}^*=\frac{d}{dt}|v|$ in the relation \eqref{astar} that defines the desired acceleration $\bm a^*$. This supposes that $\frac{d}{dt}|v|$ is either estimated or measured. This choice can also be opted for when thrust control is used to stabilize the airspeed component $v_{a,1}$ instead of $|v|$.\\
A different issue concerns thrust power limitations. Indeed, while unlimited thrust power theoretically allows unlimited speed, a finite value $T_{max}$ automatically limits the aircraft speed. Moreover, when the maximal thrust is significantly smaller than the aircraft weight, it critically matters to keep the attack angle small (and under the stall value). Without this safety feature the direction $\bar{\bm \imath}$ calculated (or imposed by a pilot) without taking this limitation into consideration may yield a stall situation and an important loss of lift leading to a continuing descent and the final crash of the aircraft.\\ 
To automatically integrate this safety feature in the proposed control design let us rename the unit vectors $\{ \bar{\bm \imath}, \bar{\bm \jmath}, \bar{\bm k} \}$ calculated without taking thrust limitations into account as $\{ \bar{\bm \imath}^*, \bar{\bm \jmath}^*, \bar{\bm k}^* \}$. Let also $\alpha_{max}$ denote the desired upper bound for the attack angle. Whenever the predicted attack angle $\alpha^*:=\arcsin(\frac{\bm v_a}{|v_a|} \cdot \bar{\bm k^*})$ is larger than $\alpha_{max}$ we propose to determine new desired attitude directions for the aircraft such that the associated attack angle is equal to $\alpha_{max}$. More precisely, we propose to set
\begin{equation} \label{newibar}
\bar{\bm \imath}=rot(\alpha_{max}\bar{\bm \jmath}^*)\frac{\bm v_a}{|v_a|},~\bar{\bm \jmath}=\bar{\bm \jmath}^*,~\bar{\bm k}=\bar{\bm \imath}\times \bar{\bm \jmath}
\end{equation}
with $rot(\alpha_{max}\bar{\bm \jmath}^*)\frac{\bm v_a}{|v_a|}$ denoting the vector $\frac{\bm v_a}{|v_a|}$ rotated by the angle $\alpha_{max}$ about the unit vector $\bar{\bm \jmath}^*$. Note that this choice corresponds to the minimization of $|\bar{\bm \imath}-\bar{\bm \imath}^*|$ under the constraint $\arcsin(\frac{\bm v_a}{|v_a|} \cdot \bar{\bm k})=\alpha_{max}$. In other words the proposed procedure minimizes the modifications brought to the calculated desired directions under the constraint of keeping the attack angle at most equal to the chosen threshold. Note also that the new attitude directions $\{ \bar{\bm \imath}, \bar{\bm \jmath}, \bar{\bm k} \}$, like $\{ \bar{\bm \imath}^*, \bar{\bm \jmath}^*, \bar{\bm k}^* \}$, do not depend on the aircraft orientation so that the stabilization of these directions via the angular velocity input $\omega$ specified in relation \eqref{omega} remains a well-posed problem. In this case the pre-compensation velocity $\bar{\bm \omega}$ is calculated with $\bm \omega_{\bar{\bm \imath}}:=\bar{\bm \imath} \times \dot{\bar{\bm \imath}}$ and, in view of \eqref{newibar}
\[
\dot{\bar{\bm \imath}}=\alpha_{max}\bm \omega_{\bar{\bm \jmath}^*}\times \bar{\bm \imath}+rot(\alpha_{max}\bar{\bm \jmath}^*){\bm \Pi}_{\frac{\bm v_a}{|v_a|}}\frac{\dot{\bm v}_a}{|v_a|}
\]
 The convergence of $\{ {\bm \imath}, {\bm \jmath}, {\bm k} \}$ to $\{ \bar{\bm \imath}, \bar{\bm \jmath}, \bar{\bm k} \}$ then ensures that the aircraft attack angle is ultimately bounded by $\alpha_{max}$.\\
When implementing this safety procedure (again imposed by a limited maximal thrust) one must further pay attention to choosing the desired speed $v^*$ larger (with some security margin) than the horizontal speed that can be sustained when the attack angle is kept equal to $\alpha_{max}$. 
\vspace{-0.2cm}
\subsection{From desired body angular velocity to control-surfaces angles velocities} \label{control_surfaces}
\vspace{-0.2cm}
Consider a fixed-wing aircraft equipped with standard roll-pitch-yaw control surfaces. Let $\delta=[\delta_1,\delta_2,\delta_3]^{\top}$ denote the vector of control-surfaces angles that have to be calculated from the desired body angular velocity $\omega^*$ determined previously. The preliminary step relies on the determination/estimation of the matrix-valued function $A(v_a,\omega)=:\frac{\partial \Gamma}{\partial \delta}_{|\delta=0}$ that relates angles deviations to torque production for small angles. This matrix depends in particular on the placement of the control surfaces with respect to the aircraft CoM and on their dimensions.  Typically, along classical trim trajectories and nominal air velocities involving small attack and side-slip angles, $A(v_a,\omega)$ can be approximated by $|v_a|^2\bar{A}$, with $\bar{A}$ a constant matrix. Let us thus assume that ${A}$, or at least an estimate of this matrix-valued function, is known. Since $J\dot{\omega}\approx -S(\omega)J\omega+\Gamma$ one deduces that
a torque equal to $\Gamma^{\star}=S(\omega)J\omega^*-k_{\gamma}J(\omega-\omega^*)$, with $k_{\gamma}$ a large positive gain, is effective to make $\omega$ track $\omega^*$. This in turn suggests to make $\delta$ track $\delta^{\star}=:{A}^{-1}\Gamma^{\star}$ by setting $\dot{\delta}=u_{\delta}$ with $u_{\delta}=:-k_{\delta}(\delta-\delta^{\star})$ and $k_{\delta}$ a positive gain. However, control surfaces angular velocities, as well as angles, are usually bounded. Let $\bar{u}_{\delta,i}$ ($i=1,2,3$) denote the saturated value of $u_{\delta,i}$, and $\Delta_{\delta_i}$ the maximal value of $\delta_i$. A possible driving law for each control surface angle that respects these bounds is then
\[
\dot{\delta}_i=-\bar{k}_{\delta}\delta_i+\bar{k}_{\delta}\bar{\sat}^{\Delta_{\delta_i}}(\delta_i+\bar{u}_{\delta,i}/\bar{k}_{\delta})~,~i=1,2,3
\]
with $\bar{k}_{\delta}$ denoting a large positive gain.
\vspace{-0.2cm}
\section{Simulations} \label{simulations}
\vspace{-0.2cm}
\subsection{Hardware-in-the-loop}
\vspace{-0.2cm}
The object of this section is to test and validate via realistic simulation the path following control approach reported in the previous sections. A particularity of the proposed simulations is that they are {\em hardware-in-the-loop} implemented in the sense that the control algorithms are implemented on a piece of hardware that could equip a physical UAV aircraft. The only difference with a true experimentation is that this hardware is connected to a computer simulated model of an aircraft that closely mimics the dynamics of a physical aircraft. This type of simulation is widely recognized for its convenience and soundness.
\vspace{-0.2cm}
\subsubsection{Aircraft simulator}
\vspace{-0.2cm}
We use the {\em X-plane 10} software, a Laminar Research product, as a flight simulator (http://www.x-plane.com/desktop/how-x-plane-works/). X-plane implements an aerodynamic model based on the so-called blade element theory. 
For the present simulations we have used a model of a small RC aircraft with the following specifications:
\vspace{-0.2cm}
\begin{itemize}
  \item Wingspan: $1.5 m$
  \item Wing surface: $0.36 m^2$
  \item Fuselage length: $1 m$
  \item Weight: $2 kg$
\end{itemize}
  \vspace{-0.2cm}
\subsubsection{Controller hardware and software}
\vspace{-0.2cm}
A commercially available "Pixhawk", equipped with a 168 MHz ARM Cortex M4F CPU and 256 KB of RAM, is used as the autopilot hardware. Our code implementation is based on the open-source PX4 flight stack that runs on top of a NuttX real-time operating system and uses the PX4 middleware \cite{px4}. This software architecture runs different threads with a modular approach and handles inter-process communications, allowing for the development and use of off-the-shelf control code. We took advantage of this possibility to replace the pre-existing position and attitude control modules by our own libraries in order to implement the control algorithms presented in the paper. However we kept the other pre-existing functionalities and, in particular, the EKF state observer.
Flight data is logged on a SD card, allowing for post-processing and analysis of this data.
\vspace{-0.2cm}
\subsubsection{Ground control station}
\vspace{-0.2cm}
We use the "Qgroundcontrol" software\footnote{http://qgroundcontrol.com/} as a control station installed on a base computer, to design missions and change parameters online during flights. In our simulation setup, Qgroundcontrol connects the base computer to the "PIXHAWK" controller via USB using the "MAVLINK" protocol. In parallel, Qgroundcontrol establishes a UDP link with the X-plane simulator to send and receive data. This dual connection allows Qgroundcontrol to perform hardware-in-the-loop simulation by allowing an indirect communication between the controller and the simulator. Simulated position and attitude of the aircraft are used to create virtual sensory measurements (GPS, IMU, barometer, pitot) that are purposefully corrupted with artificial noise and produced (sent) at a realistic rate. These measurements are processed by the included EKF state observer and are used to calculate PWM (Pulse-width-Modulation) control values. The controller sends these control values to the ground station, which in turn sends them to the simulator that uses them as set points for the airplane's propeller angular velocity and for the control surfaces angles.

\begin{figure}
\centering
 \includegraphics[height=6.5cm]{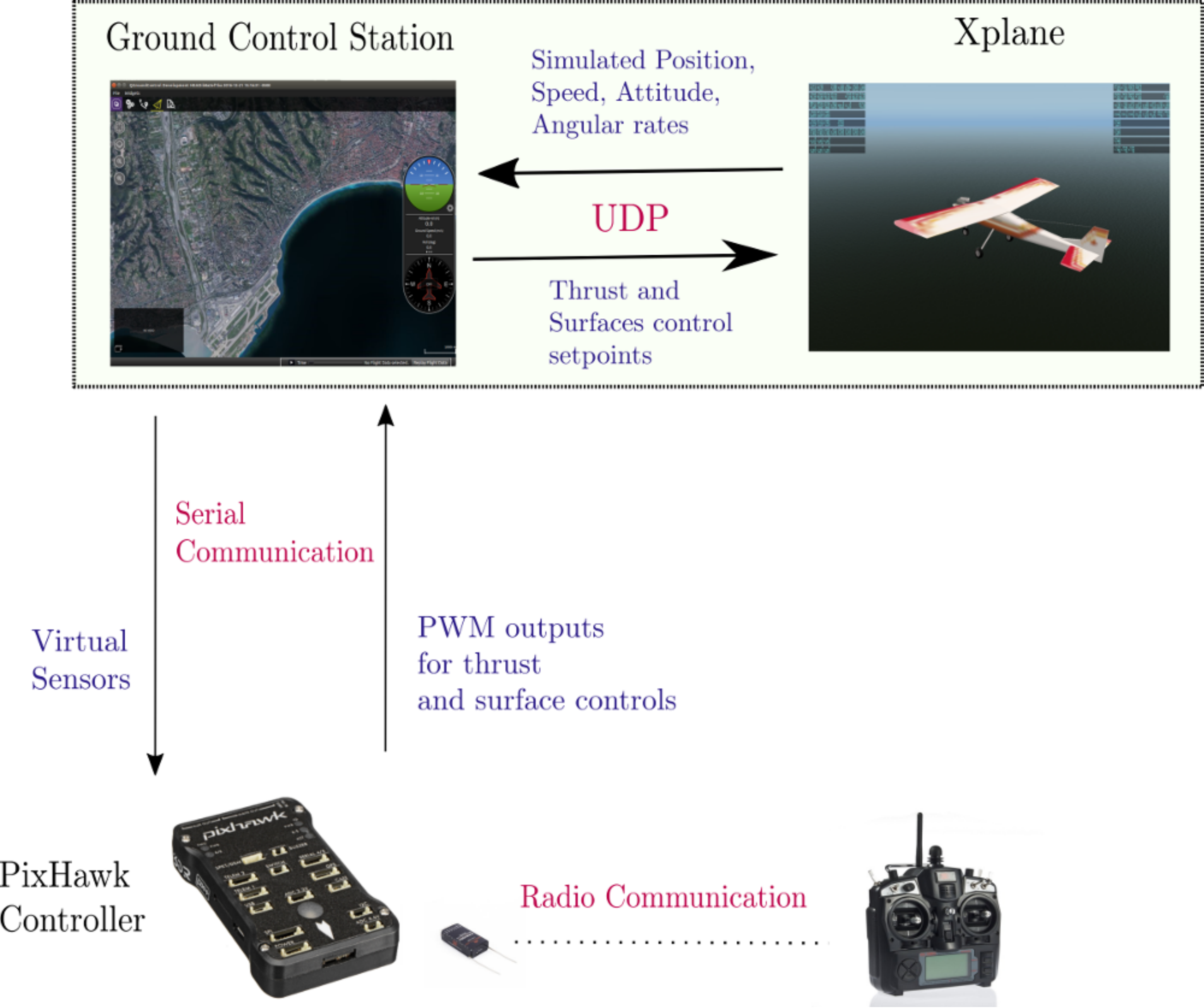}
 \caption{HITL tools}
 \label{fig:hitl}
\end{figure}
\vspace{-0.2cm}
\subsection{Simulation Results}
\begin{figure}
 \includegraphics[width=.48\textwidth]{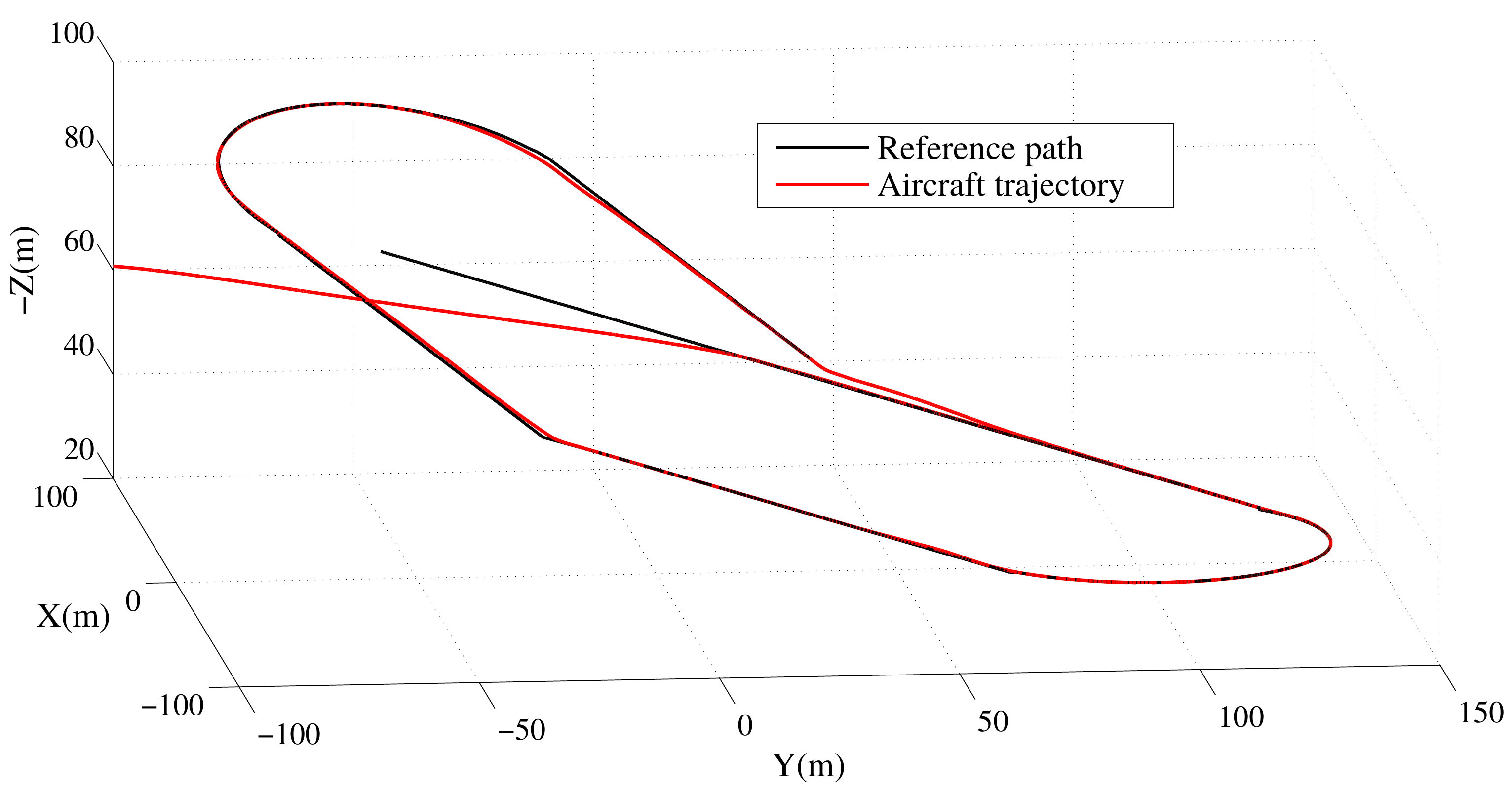}
 \caption{Aircraft trajectory and reference path}
 \label{fig:4pos3dwind}
\end{figure}

\begin{figure}
 \includegraphics[width=.48\textwidth]{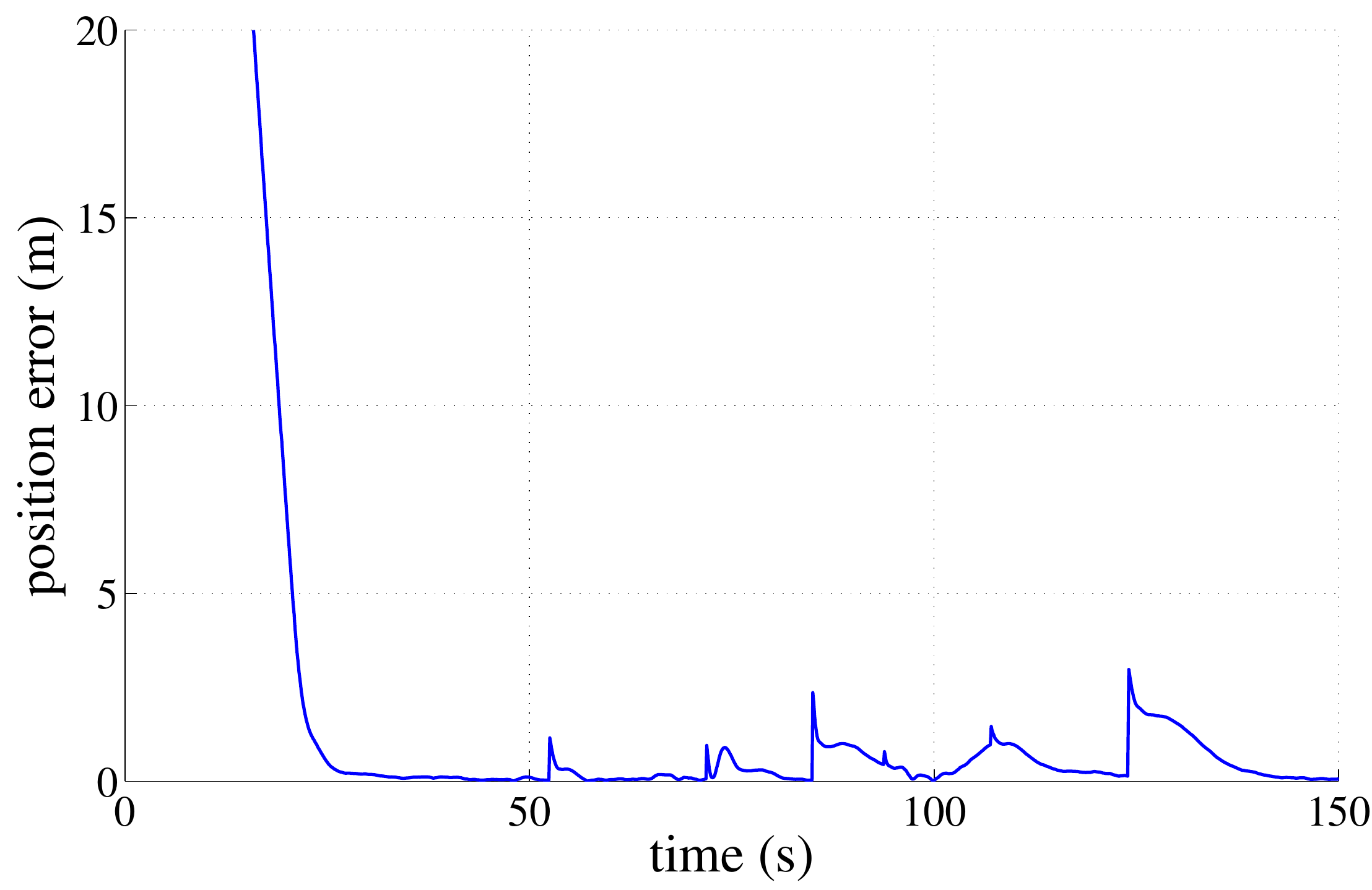}
 \caption{Position error $|y|$}
 \label{fig:4ywind}
\end{figure}

\begin{figure}[htb]
\centering
  \begin{tabular}{@{}cccc@{}}
    \includegraphics[width=.48\textwidth]{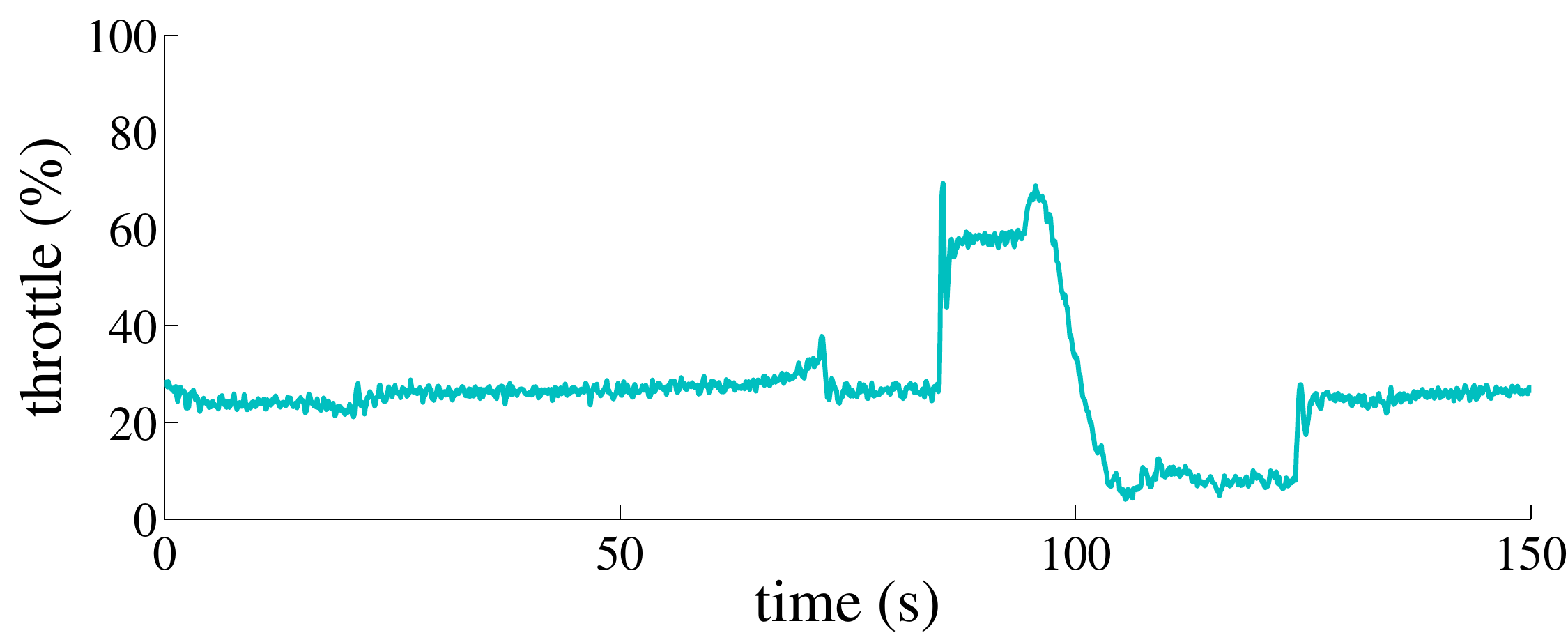} \\
    \includegraphics[width=.46\textwidth]{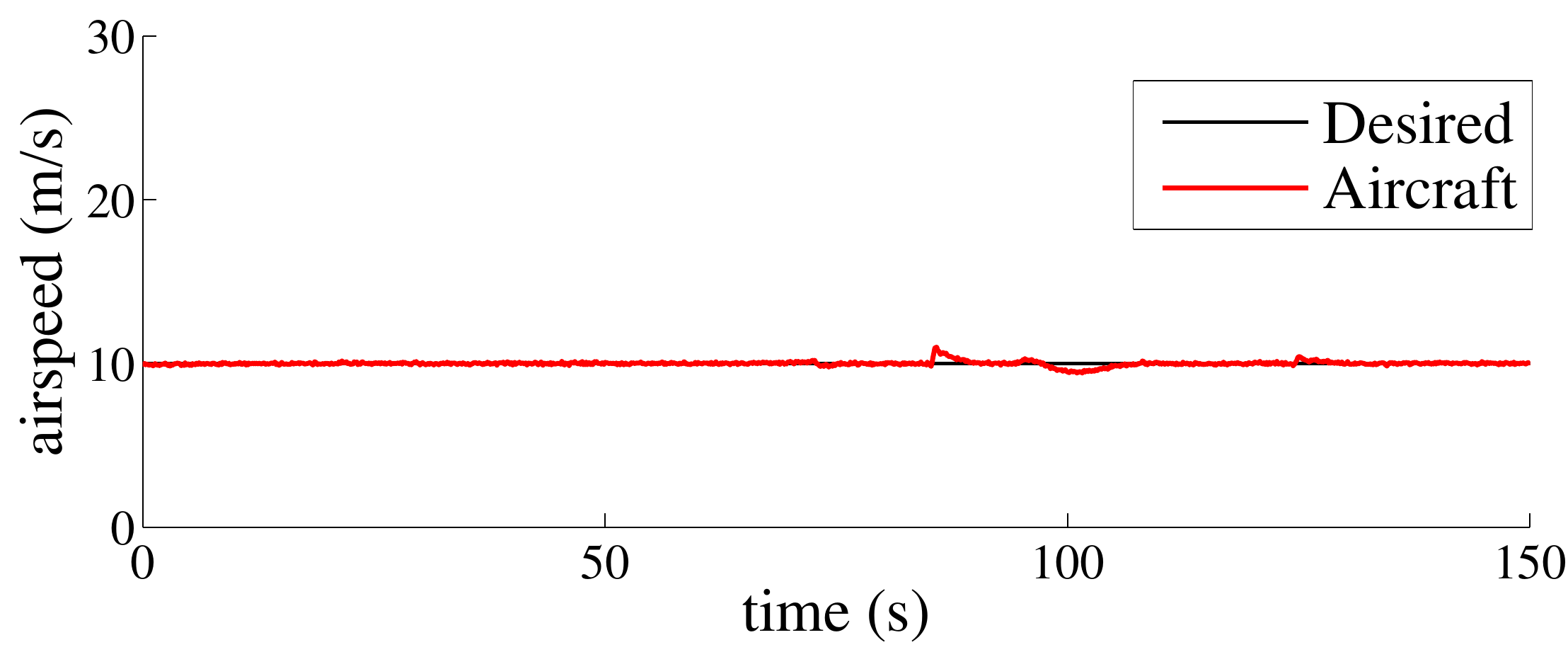}\\
    \includegraphics[width=.48\textwidth]{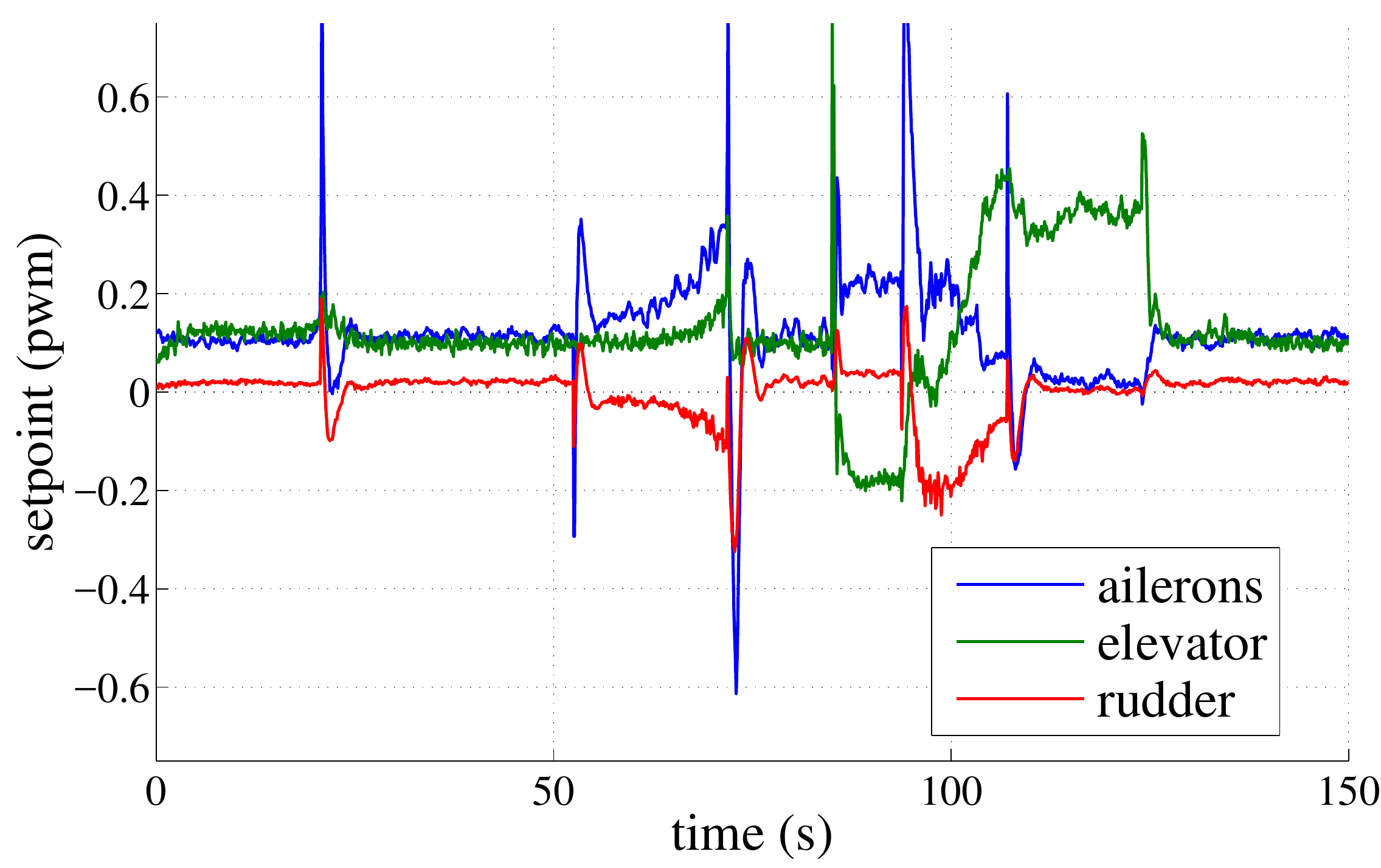}\\
    \includegraphics[width=.48\textwidth]{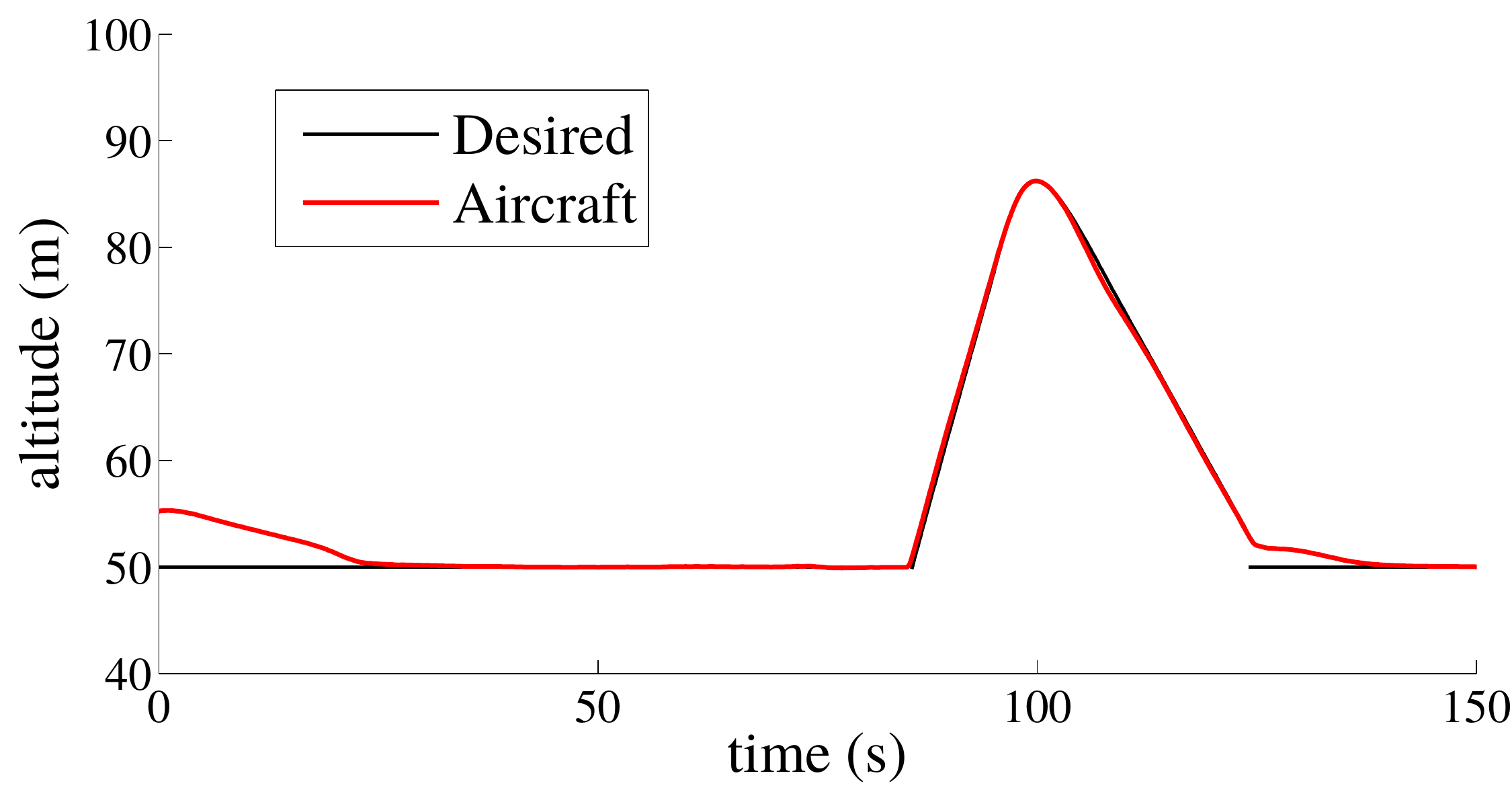}
  \end{tabular}
  \caption{3-axis aircraft: (a) desired throttle (b) airspeed $v_{a,1}$ (c) angles of control surfaces}
  \label{fig2:thrspeed}
\end{figure}

\vspace{-0.2cm}
For these simulations the aircraft inertial position $\bm p$, velocity $\bm v$, attitude $(\bm \imath, \bm \jmath, \bm k)$, and angular velocity $\omega$ are supposed to be measured (by using on-board GPS and gyro sensors, for instance) and are thus available for control computations. The air-velocity component $v_{a,1}$, measured by a Pitot tube along the aircraft longitudinal direction $\bm \imath$, is also available and the demanded airspeed for $v_{a,1}$ is $10 m/s$. For the reported simulations we have assumed that an {\em a priori} unknown wind may be present so that we can never presume that the air-velocity $\bm v_a$ is equal to the aircraft inertial velocity $\bm v$. Beside $v_{a,1}$ the other components of $\bm v_a$ involved in the control expressions can be measured and/or estimated in various manners by fusing the information provided by complementary sensors. It is also possible to derive and use an approximation of $\bm v_a$ based on the aircraft model equations \eqref{Fa}-\eqref{eq:newton0-1} and on the sole direct measurement of the component $v_{a,1}$. Let us further elaborate on this possibility. From \eqref{acceleration}, and considering that the non-specified transversal component $v_{a,2}\bm \bar{O}(\bm v_a)$ of the resultant aerodynamic force is parallel to $\bm \jmath$, one gets $\bm a.\bm k=\bar{\bm g} \cdot \bm k$. Therefore, using the expression \eqref{gbar} of $\bar{\bm g}$ one deduces that $v_{a,3}=\bm v_a \cdot \bm k=\frac{m}{\bar{c}_0|v_a|}(\bm g-\bm a) \cdot \bm k$. Approximating $|v_a|$ by $|v_{a,1}|$ and using an estimation $\hat{\bm a}$ of the aircraft acceleration, an approximation of $v_{a,3}$ is $\hat{v}_{a,3}=\frac{m}{\bar{c}_0|v_{a,1}|}(\bm g-\hat{\bm a}) \cdot \bm k$. To estimate $\hat{\bm a}$ one may use on-board accelerometers that measure the gravity acceleration minus the body acceleration, i.e. $\bm {acc}=\bm g - \bm a$ so that $\bm a= \bm g - \bm {acc} \ (= \hat{\bm a})$. Another  possibility consists in estimating $\bm a$ from the measurement of the inertial velocity $\bm v$ obtained, for instance, with a GPS. Further assuming that $\bm v$ is almost constant in the body frame yields the crude estimation $\hat{\bm a}= \bm \omega \times \bm v$. An even cruder estimation is obtained by assuming that the aircraft acceleration is small, i.e. $\hat{\bm a} \simeq 0$. Concerning the sideslip velocity component $v_{a,2}$, we assume that it is nominally small by virtue of the strong energy dissipation along the aircraft transversal direction that results from the existence of a tail rudder and, to a smaller extent, from the oblong shape of the body aircraft (this dissipation is actively re-enforced by the balanced flight control policy). The resulting approximation of $\bm v_a$ is then $\hat{\bm v}_a=v_{a,1}\bm \imath+\hat{v}_{a,3} \bm k$. This solution seems convenient for scale-model UAVs that are not equipped with dedicated sensors that measure $\bm v_a$ in all three directions. We have used it with $\hat{\bm a} = 0$ to test the control robustness.\\
The aerodynamic coefficients used for the control calculations are: $c_0=0.006$, and $c_1=0.5$. By application of relations \eqref{gr} and \eqref{vsink}, these coefficients correspond to a glide rate equal to $6.5$ and a gliding speed equal to $15.9 m/s$.\\
The parameters used in the control laws are:
\vspace{-0.2cm}
\begin{itemize}
    \item Guidance: $k_1=1$, $\mu=0.5$, $d_1=1$, $d_2=0.5$
    \item Velocity/airspeed control: $k_{T,1}=1.8$, $k_{T,2}=k_{T,1}/2$
    \item Heading stabilization: $k_{h,1}=1.4$, $k_{h,2}=0.49$, $\Delta_z=0.5$, $k_z=10$
    \item Attitude control: $k_{\omega}=7$
  \end{itemize}
\vspace{-0.3cm}
As explained in section \ref{control_surfaces}, control surfaces angles are calculated according to
$\delta^\ast=-\frac{k_\gamma}{|v_a|^2} \bar{A}^{-1}J(w-w^\ast)=-\frac{1}{|v_a|^2}diag([45,60,45])(w-w^\ast)$. The term $S(w)J w^\ast$ is purposefully neglected because its contribution is not important compared to other terms. The actuators angular velocities are also limited to $1rad/s$ by simply monitoring the difference between computed consecutive angle values.\\
The chosen closed reference curve (see Fig. \ref{fig:4pos3dwind}) consists, for the first part, of a horizontal segment connected to a horizontal half-circle of radius equal to $50 m$, followed by another horizontal segment. The second part of the reference path is similar to the first one except that it is inclined with an angle of $15^\circ$ w.r.t. the horizontal plane.
Just recall here that an inclined circular path does not qualify as a trim trajectory.\\
A steady ({\em a priori} unknown) wind of intensity $|v_w|=3 m/s$ blowing from the South (corresponding to the X-axis in Fig. \ref{fig:4pos3dwind}) has been added to the X-plane scenario. 
Simulations results reported in Fig. \ref{fig:4pos3dwind} and Fig. \ref{fig:4ywind} show a steady position error well under the accepted norm of a wingspan, despite approximations involved in the estimation of the air-velocity $\bm v_a$, imperfections of the model used for the control design, and the noisy state estimates produced by the embedded EKF observers. The observable discontinuities on the figure correspond to switching time-instants, followed by short transient convergence phases, when the position error is measured w.r.t. a new part of the path.
Fig. \ref{fig2:thrspeed} shows the time evolution of various variables.
\vspace{-0.2cm}
\section{Concluding remarks}\label{conclusion}
\vspace{-0.2cm}
The present paper reports on a novel nonlinear control approach to the path following problem for scale-model airplanes. This approach departs from other ones by developing a unified framework for kinematic guidance and dynamic control. The approach builds upon a simple analytic model of aerodynamic forces acting on the vehicle, and the resulting nonlinear controllers are designed to operate in a large spectrum of operating conditions. In particular they overcome the limitations associated with classical methods based on linearization of aircraft dynamics equations along so-called trim trajectories. Reported realistic hardware-in-the-loop simulations illustrate the proposed control methodology, and its robustness w.r.t modeling, measurements, and estimation approximations. The next logical stage of this study will involve intensive experimentation on different scale-model airplanes.
\vspace{-0.3cm}
\appendix
\section{Proof of Proposition \ref{speed_control} }
\vspace{-0.3cm}
Consider the following dynamic system on $\RR^n \times \RR^n$
\begin{equation} \label{S}
\left\{ 
\begin{array}{lll}
\dot{x}&=&-k_1x-k_2\alpha_x y+\mu\\
\dot{y}&=&k_2k_3\big(-y+\alpha_x(y+x/k_3)\big)
\end{array}
\right.
\end{equation}
with
\begin{itemize}
\item $k_3 \in \RR^+-\{0\}$;
\item $|\mu(x)|\leq k_{\mu}(|x|+c)^q$, with $k_{\mu} \in \RR^+$, $c\in \RR^+$, $q\in \RR^+-\{0\}$; 
\item $k_1(x)=k_{11}+k_{12}|x|^p$, with $p\in \RR^+-\{0\}$, $p>max(0,q-1)$, $k_{11}\in \RR^+-\{0\}$, $k_{12}\in \RR^+-\{0\}$;
\item $k_2(x)>0$, $\forall x$, and such that $0<r_1 \leq \frac{k_1(x)}{k_2(x)} \leq r_2<+\infty$;
\item $\alpha_x(x,y)=\alpha^{\Delta_{y}}(|y+x/k_3|)$, with $\Delta_{y} \in \RR^+-\{0\}$.
\end{itemize}
By definition of the function $\alpha^{\Delta_{y}}$ (see Section \ref{sec-back}), $\alpha_x(y+x/k_3)=\bar{\sat}^{\Delta_y}(y+x/k_3)$ ($\leq \Delta_y$). The vector $y$ is a bounded integral of $x$, and the gain $k_3$ is chosen large enough to ensure fast de-saturation. The system \eqref{S} would be a classical second order differential linear system, yielding the exponential stability of the point $(x,y)=(0,0)$, if $(\alpha_x,k_{12},k_{\mu})$ were equal to $(1,0,0)$ and $r_1$ were equal to $r_2$. The role of the positive gain $k_{12}$, which may be chosen arbitrarily small, is to ensure the boundedness of $|x|$ when a non-zero perturbation $\mu$ is present.
\begin{lemma} \label{lemma}
The solutions $(x(t),y(t))$ to the system \eqref{S} verify the following properties:
\begin{enumerate}
\item $|y(t)|$ is bounded and ultimately bounded by $\Delta_y$.
\item $|x(t)|$ is bounded.
\item solutions are complete (exist for $t\in [0,+\infty)$).
\item If $\mu \equiv 0$ then ${\mathcal L}_{\epsilon}(x,y):=0.5 (|x|^2+|y|^2)+\epsilon x^{\top}y$, with $\epsilon$ a small enough positive number, is a Lyapunov function for the unperturbed system, and $(x,y)=(0,0)$ is globally exponentially stable.
\item If $\mu(x(t))$ converges exponentially to zero then $(x(t),y(t))$ converges exponentially to $(0,0)$.
\end{enumerate}
\end{lemma}
\vspace{-0.2cm}
\noindent {\bf Proof}:\\
{\em (1)}: From the second equation of \eqref{S}
\begin{equation} \label{AA1}
\begin{array}{lll}
0.5\frac{d}{dt} |y|^2& = & -k_2k_3\big(|y|^2-y^{\top}\bar{\sat}^{\Delta_y}(y+x/k_3)\big)\\
~ & \leq & -k_2k_3|y|(|y|-\Delta_y)
\end{array}
\end{equation}
Therefore $|y(t)|$ decreases when $|y(t)|>\Delta_y$, and the stated property follows immediately.\\
{\em (2)}: Define ${\mathcal L}_{\epsilon}(x,y):=0.5 (|x|^2+|y|^2)+\epsilon x^{\top}y$. This is a positive quadratic form provided that $0<\epsilon<1$. Along any solution to the system one verifies that
\begin{equation} \label{AA2}
\begin{array}{lll}
\dot{{\mathcal L}}_{\epsilon}&=&-(k_1-\epsilon k_2\alpha_x)|x|^2-k_2\big(k_3(1-\alpha_x)+\epsilon \alpha_1\big)|y|^2\\
~&~&-\epsilon \big(k_1+k_2k_3(1-\alpha_x)\big)x^{\top}y+(x+\epsilon y)^{\top}\mu
\end{array}
\end{equation}
In particular 
\begin{equation} \label{AA3}
\begin{array}{lll}
\dot{\mathcal L}_{0}&=&-k_1|x|^2-k_2k_3(1-\alpha_x)|y|^2+x^{\top}\mu\\
~& \leq & -k_{12}|x|^{p+2}+k_{\mu}|x|(|x|+c)^q
\end{array}
\end{equation}
Since $p+1>q$ by assumption, this latter relation shows that ${\mathcal L}_{0}(t)$ decreases when $|x(t)|$ is larger than some number $k_x$. Therefore, using the fact that $|y(t)|$ is bounded uniformly w.r.t. to the initial condition $y(0)$, one deduces that $|x(t)|$ is also bounded uniformly w.r.t the initial condition $x(0)$. Furthermore, since $|y(t)|$ is ultimately bounded by $\Delta_y$, $|x(t)|$ is ultimately bounded by $\sqrt{k_x^2+\Delta_y^2}$.\\
{\em (3)}: This property is just a consequence of the previous two properties.\\
{\em (4)}: In view of \eqref{AA2}, if $\mu\equiv 0$ and $\epsilon<k_3$ then
\begin{equation} \label{AA4}
\begin{array}{lll}
\dot{{\mathcal L}}_{\epsilon}&=&-(k_1-\epsilon k_2\alpha_x)|x|^2-k_2\big(k_3(1-\alpha_x)+\epsilon \alpha_1\big)|y|^2\\
~&~&-\epsilon \big(k_1+k_2k_3(1-\alpha_x)\big)x^{\top}y\\
~&\leq & -k_2\big((\frac{k_1}{k_2}-\epsilon)|x|^2+\epsilon|y|^2-\epsilon \big(\frac{k_1}{k_2}+k_3)|x||y|\big)\\
\end{array}
\end{equation}
From this inequality one deduces the existence of two positive numbers $\epsilon_0<1$ and $k_l$ such that $\dot{{\mathcal L}}_{\epsilon_0}\leq -k_l{\mathcal L}_{\epsilon_0}$.
The stated exponential stability property follows from the fact that ${\mathcal L}_{\epsilon_0}$ is a positive quadratic form.\\
{\em (5)}: A vanishing perturbation $\mu$ such that $|\mu(t)|\leq {\mu}_M\exp{(-k_{\mu}t)}$, when $t\geq t_0$, yields
\[
\begin{array}{lll}
\dot{\mathcal L}_{\epsilon_0}&\leq& -k_l{\mathcal L}_{\epsilon_0}+|x+\epsilon_0 y||\mu|\\
~&\leq&-k_l{\mathcal L}_{\epsilon_0}+2{\mu}_M\sqrt{{\mathcal L}_{\epsilon_0}}\exp{(-k_{\mu}t)}
\end{array}
\]
when $t\geq t_0$, and thus
\[
\dot{\mathcal L}^{0.5}_{\epsilon_0} \leq 0.5(-k_l{\mathcal L}^{0.5}_{\epsilon_0}+2{\mu}_M\exp{(-k_{\mu}t)})
\]
when $t\geq t_0$. The exponential convergence of ${\mathcal L}_{\epsilon_0}(t)$, and thus of $|x(t)|$ and $|y(t)|$, to zero follows immediately.
$\square$\\~\\
Proposition \ref{speed_control} is a direct application of the technical Lemma \ref{lemma}, after setting $x:=e_v$, $y:=I_{e_v}$, $k_1:=k_{T,1}$, $k_2:=k_{T,2}$, $k_3:=k_{T,3}$, $\alpha_x=\alpha_1$, and $\mu:=v_{a,2}\bm \bar{O}(\bm v_a) \cdot \bm h$.
\vspace{-0.3cm}
\section{Proof of Proposition \ref{stabilisation}}
\vspace{-0.3cm}
Define $\bar{\bm o}:=|v|\Piu \bm o$. 
The time-variation of $\bmtildepF$ is given by
\begin{equation} \label{deriv_ptildeF}
\begin{array}{lll}
\dotbmtildepF&=&\Piu \bm v= |v|\Piu (\bm h^*+\bm o)\\
~&=& |v|\Piu \sin(\theta_h)\bm l+\bar{\bm o}\\
~&=& |v| |\bar{y}| \bm l+\bar{\bm o}\\
~&=& -|v|(\bar{y}_1\barbmu+\bar{y}_2\barbarbmu)+\bar{\bm o}
\end{array}
\end{equation}
Note that
$\int_0^t |\bar{\bm o}(s)|ds \leq \sup(|v|)\int_0^t |\bm o(s)|ds$ and that this integral is thus bounded.
Because $\dotbmtildepF=\dot{y}_1\barbmu+\dot{y}_2\barbarbmu$, it comes that $\dot{y}=-|v|\bar{y}=-k_1D\bar{\sat}^{\Delta_{h}}(y)+\bar{o}$, with $\bar{o}$ denoting the vector of coordinates of $\bar{\bm o}$ along the unit vectors $\barbmu$ and $\barbarbmu$. The announced ultimate upper bounds of $|\dot{y}(t)|$ and $|\dot{y}_i(t)|$ ($i=1,2$) follow directly from this latter equality and the convergence of $|\bar{o}|$ to zero imposed by the boundedness of the integral of this term.
The convergence of $y$ and $\dot{y}$ to zero then follows from re-writing the previous equality as $\dot{y}_i=-k_1d_i\alpha^{\Delta_h}(|y|)y_i+\bar{o}_i$ ($i=1,2$), so that $\frac{d}{dt}|y_i|\leq-k_1d_i\alpha^{\Delta_h}(|y|)|y_i|+|\bar{o}_i|$. The rate of convergence is ultimately exponential when $|\bm o|$, and thus $|\bar{\bm o}|$, themselves converge ultimately exponentially to zero. The local exponential stability of $y=0$ in the sense of Lyapunov when $\bm o \equiv \bm 0$ is inherited from the non-saturated equation $\dot{y}=-k_1Dy$ that holds in the first approximation when $|y|$ is small. Finally, since $|y|$ and thus $|\bar{y}|$ tend to zero, $\theta_h$ converges to zero and, in view of \eqref{hstar}, $\bm h$ ($=\bm h^*$) converges to $sign_{v_u}\bm u$.
\vspace{-0.3cm}
\section{Proof of Proposition \ref{hstabilisation}}
\vspace{-0.3cm}
Forming the time-derivative of the positive function ${\mathcal V}_0:=(1-\bm h.\bm h^*)+0.5k_{h,2}|\bm z|^2$ yields
\begin{equation} \label{deriv_V0}
\dot{\mathcal V}_0=-k_{h,1}|\tilde{\bm h}|^2-k_{h,2}k_z(1-\alpha_h)|\bm z|^2-\bm o\cdot \tilde{\bm h}
\end{equation}
\noindent {\bf case 1}: $\forall t:~\bm o(t)= \bm 0$.\\
In this case $\dot{\mathcal V}_0 \leq 0$, and $\dot{\mathcal V}_0=0$ iff $(\bm h,\bm z)=(\bm h^*,\bm 0)$ or  $(\bm h,\bm z)=(-\bm h^*,\bm 0)$. It is simple to verify that these two points, for which ${\mathcal V}_0$ is stationary, are indeed equilibria of the system \eqref{omegh}, \eqref{saturated_integral}. 
Because ${\mathcal V}_0$ is non-increasing, and $\alpha_h(\bm 0,\bm z)=1$ only if $\bm z=\bm 0$, the convergence of $|\tilde{h}|$ and $|z|$ to zero follows. This in turn implies that $(\bm h,z)$ converges to one of the system's equilibria. Because $1-\bm h(t) \cdot \bm h^*\leq {\mathcal V}_0(t)\leq {\mathcal V}_0(0)$ is always smaller than two when $\bm h(0) \neq -\bm h^*$, $\bm h$ cannot converge to $-\bm h^*$, and thus necessarily converges to the desired equilibrium.\\
The local exponential stability of the equilibrium  $(\bm h,\bm z)=(\bm h^*,\bm 0)$ follows from an adaptation of the technical Lemma \ref{lemma} reported in the Appendix consisting in showing that, in the neighborhood of this equilibrium, there exist two positive numbers $\epsilon$ and $k_h$ such that i) the approximation, at the second order in a system of Cartesian coordinates of $\tilde{\bm h}$ and $\bm z$, of
${\mathcal V}_{\epsilon}:=(1-\bm h.\bm h^*)+0.5k_{h,2}|z|^2+\epsilon \bm z \cdot \tilde{\bm h}$ about the equilibrium $(\bm h,\bm z)=(\bm h^*,\bm 0)$ is a positive quadratic form, and ii) $\dot{\mathcal V}_{\epsilon}\leq -k_h{\mathcal V}_{\epsilon}$. Another way of establishing the stability, or instability, properties of the system's equilibria consists in studying linear approximations of the system about these equilibria. Consider a frame centered on the aircraft CoM and rotating with the angular velocity $\bm \omega_{h^*}$, and let $x_1$ (resp. $x_2$) denote the two-dimensional vector of Cartesian coordinates (in this frame) of the projection of $\tilde{\bm h}$ (resp. $\bm z$) onto the plane orthogonal to $\bm h^*$. One verifies that, in the first approximation about $(\tilde{\bm h}, \bm z)=(\bm 0,\bm 0)$, the variations of $x_1$ and $x_2$ satisfy the equations of the linear system
\begin{equation} \label{linearized_system}
 \left\{\begin{array}{lll}
\dot{x}_1&=&\mp (k_{h,1}x_1+k_{h,2}x_2)\\
\dot{x}_2&=& x_1
\end{array} \right.
\end{equation}
with the sign in the right-hand side of the first equality depending on the chosen equilibrium, i.e. $(\bm h,\bm z)=(\bm h^*,\bm 0)$ or $(\bm h,\bm z)=(-\bm h^*,\bm 0)$. The minus sign goes with the first equilibrium, and the plus sign with the second one. The characteristic polynomial associated with the first (resp. second) one is $(\lambda^2+k_{h,1}\lambda+k_{h,2})^2=0$ (resp.   
$(\lambda^2-k_{h,1}\lambda-k_{h,2})^2=0$). All poles of the first system have negative real parts, whereas two poles of the second system are real positive. This in turn proves that the equilibrium $(\bm h,\bm z)=(\bm h^*,\bm 0)$ is exponentially stable, and that the other equilibrium is (exponentially) unstable.\\~\\
\noindent {\bf case 2}: $\exists t:~\bm o(t)\neq \bm 0$.\\
Let us first establish that $(\tilde{\bm h},\bm z)$ converges to $(\bm 0,\bm 0)$. 
The assumed boundedness of $\int_0^t |\bm o(s)|ds$ implies that $\int_0^t \bm o(s) \cdot \tilde{\bm h}(s)ds$ is also bounded. Since ${\mathcal V}_0$ is positive and uniformly bounded from above, $|{\mathcal V}_0(t)-{\mathcal V}_0(0)+\int_0^t \bm o(s) \cdot \tilde{\bm h}(s)ds|=k_{h,1}\int_0^t |\tilde{\bm h}(s)|^2ds+k_{h,2}k_z\int_0^t (1-\alpha_1(s))|\bm z(s)|^2ds$ is also uniformly bounded. Therefore the integrals $\int_0^t |\tilde{\bm h}(s)|^2ds$ and $\int_0^t (1-\alpha_1(s))|\bm z(s)|^2ds$ are uniformly bounded. Because the time-derivative of $|\tilde{\bm h}|$ is bounded, the first of these integrals would diverge if $|\tilde{\bm h}|$ did not converge to zero. The same reasoning applies to the second integral and leads to the convergence of $(1-\alpha_1)|\bm z|^2$ to zero. This term also converges to $\big(1-\alpha^{\Delta_z}(|\bm z|)\big)|\bm z|^2=\big(|\bm z|-\bar{\sat}^{\Delta_z}(|\bm z|)\big|\bm z|$ when $\tilde{\bm h}$ converges to zero. Because $\bar{\sat}^{\Delta_z}(|\bm z|)<|\bm z|$ when $\bm z \neq \bm 0$, the convergence of this latter term to zero in turn implies the convergence of $|\bm z|$ to zero.\\
The convergence of $(\tilde{\bm h},\bm z)$ to $(\bm 0,\bm 0)$ in turn implies that $({\bm h},\bm z)$ converges either to $(\bm h^*,\bm 0)$ or to $(-\bm h^*,\bm 0)$. Therefore, non-convergence to the unstable point $(-\bm h^*,\bm 0)$ implies convergence to the desired stable point $(\bm h^*,\bm 0)$. The ultimate exponential rate of convergence when $|\bm o|$ converges to zero exponentially follows from the fact that an additive, exponentially vanishing, perturbation applied to a system whose origin is exponentially stable does not prevent the solutions to this system from converging to zero exponentially. See also the proof of the point (5) of the technical lemma \ref{lemma}.
\vspace{-0.3cm}
\section{Proof of Proposition \ref{framestabilisation}}
\vspace{-0.3cm}
Let $\tilde{\theta}$ denote the angle between the frames $\bar{\mathcal B}$ and ${\mathcal B}$, and $\bm \nu$ the unitary vector along the axis of rotation. The trace of the rotation matrix between these frames is $\mbox{tr}(\tilde{R})=(\bm \imath \cdot \bar{\bm \imath}+\bm \jmath \cdot \bar{\bm \jmath}+\bm k \cdot \bar{\bm k})$. Differentiating w.r.t. time the classical identity $\mbox{tr}(\tilde{R})=1+2\cos(\tilde{\theta})$, which is equivalent to
\[
4 \sin^2(\tilde{\theta}/2)=3-\mbox{tr}(\tilde{R})
\]
then yields
\[
4 \frac{d}{dt}\sin^2(\tilde{\theta}/2)=(\bar{\bm \omega}-\bm \omega) \cdot \big((\bm \imath \times \bar{\bm \imath})+(\bm \jmath \times \bar{\bm \jmath})+(\bm k \times \bar{\bm k})\big)
\]
with
\[
2\sin(\tilde{\theta})\bm \nu=(\bm \imath \times \bar{\bm \imath})+(\bm \jmath \times \bar{\bm \jmath})+(\bm k \times \bar{\bm k})
\]
Therefore, applying the angular velocity \eqref{omega} yields
\[
\frac{d}{dt}\sin^2(\tilde{\theta}/2)=-k_{\omega}\sin^2(\tilde{\theta})=-4k_{\omega}\sin^2(\tilde{\theta}/2)\cos^2(\tilde{\theta}/2)
\]
so that
\[
\frac{d}{dt}\tan^2(\tilde{\theta}/2) = \frac{\frac{d}{dt}\sin^2(\tilde{\theta}/2)}{\cos^4(\tilde{\theta}/2)}=-4k_{\omega}\tan^2(\tilde{\theta}/2)
\]
In other words the time-derivative of the positive function
${\mathcal V}=\tan^2 (\frac{\tilde{\theta}}{2}))$ is equal to $\dot{\mathcal V}=-4k_{\omega}(t){\mathcal V}$ $(\leq 0)$. Almost global exponential stability of $\tilde{\theta}=0$ follows immediately. Stability is not global because orientations such that $\tilde{\theta}=\pi$ are (unstable) equilibria.
\vspace{-0.3cm}
\section{Proof of Theorem \ref{hstabilisation2}}
\vspace{-0.3cm}
The convergence of $|v|$ to $v^*$ and the boundedness of $|v|$ when $|\bm \imath \cdot \bm v|$ remains larger than some positive number was established in Proposition \ref{speed_control}.
The convergence of the aircraft frame to $\bar{\mathcal B}$ when $\tilde{\theta}(0) \neq \pi$ was established in Proposition \ref{framestabilisation}. Therefore, since $|v|$, and thus $|v_a|$, are bounded,
$v_{a,2}=\bm v_a \cdot \bm \jmath=\bm v_a \cdot (\bm \jmath-\bar{\bm \jmath})$ and the sideslip angle converge to zero.
From the convergence of $\bm \imath$ to $\bar{\bm \imath}$ one deduces from relations \eqref{bmi} and \eqref{ibar} that $\xi:=\frac{{\bm a} -\bm{\bar{g}}}{|{\bm a} -\bm{\bar{g}}|}-\frac{{\bm a^*} -\bm{\bar{g}}}{|{\bm a^*} -\bm{\bar{g}}|}$ converges to zero.
The convergence of $|v|-v^*$ to zero entails the convergence of $\frac{d}{dt}|v|-\dot{v}^*$ to zero. Since $\bm a-\bm a^*=(\frac{d}{dt}|v|-\dot{v}^*)\bm h+|v|(\dot{\bm h}-\bar{\bm \omega}_h \times \bm h)$ one then deduces that $(\bm a-\bm a^*) \cdot \bm h$ converges to zero. This latter property combined with the convergence of $\xi$ to zero in turn implies that $((\bm a -\bm{\bar{g}}) \cdot \bm h)(\frac{1}{|{\bm a} -\bm{\bar{g}}|}-\frac{1}{|{\bm a^*} -\bm{\bar{g}}|})$ converges to zero. Since $|(\bm a -\bm{\bar{g}}) \cdot \bm h|$ is, by assumption, always larger than some positive number,
$(|{\bm a} -\bm{\bar{g}}|-|{\bm a^*} -\bm{\bar{g}}|)$ converges to zero, and so does
$(\bm a-\bm a^*)$. Therefore $(\dot{\bm h}-\bar{\bm \omega}_h \times \bm h)$ also converges to zero. Because $\bm \omega_h:=\bm h \times \dot{\bm h}$ and $\bm \omega_h \cdot \bm h=0$ one then infers that $\bm \omega_h$ converges to $\Pi_{\bm h}\bar{\bm \omega}_h$. All convergence rates evoked so far --the rate of convergence of $\bm \omega_h$ to $\Pi_{\bm h}\bar{\bm \omega}_h$ in particular-- are ultimately exponential. Therefore $\bm \omega_h=\Pi_{\bm h}\bar{\bm \omega}_h+\bm o$, with $|\bm o|$ vanishing ultimately exponentially, and thus such that the integral $\int_0^t|\bm o(s)|ds$ is bounded. By application of Proposition 
\ref{hstabilisation}, if $(\bm h,\bm z)$ does not converge to the asymptotically unstable point $(-\bm h^*,\bm 0)$, then it converges to $(\bm h^*,\bm 0)$ with a rate of convergence also ultimately exponential. In this latter case, by application of Proposition \ref{stabilisation}, the ultimate exponential convergence of the path following error $y$ to zero, and of $\bm h$ to $sign_{v_u}\bm u$, follows.
\vspace{-0.3cm}
\section{Proof of Theorem \ref{local_stability}}
\vspace{-0.3cm}
Let $\tilde{r} \in \RR^3$ denote a local parametrization of the orientation error between the frames $\mathcal{B}$ and $\bar{\mathcal{B}}$, and $h_e \in \RR^3$ the vector of coordinates of $\bm h -\bm h^*$ in the inertial frame. Recall that $(0,y^{\top})^{\top}$ is the vector of coordinates of the position error $\bm p-\bm q$ in the frame $\mathcal{F}$.
Define $x_1:=\tilde{r}$, $x_2:=(e_v,I_{e_v})^{\top}$, $x_3:=(h_e^{\top},z^{\top})^{\top}$, $x_4:=(0,y^{\top})^{\top}$, and $x:=(x_1^{\top},x_2^{\top},x_3^{\top},x_4^{\top})^{\top} \in \RR^{14}$. Consider the error system
\[
\dot{x}=f(t,x)
\]
whose origin $x=0$ is an equilibrium. It is clear that the (local) exponential stability of this equilibrium is equivalent to the (local) exponential stability property stated in Theorem \ref{local_stability}. In view of Propositions  \ref{speed_control}-\ref{framestabilisation} and their proofs, in the neighborhood of $x=0$ this system is an interconnection of sub-systems in the form
\[
\begin{array}{l}
\dot{x}_1=f_1(t,x_1)\\
\dot{x}_i=f_i(t,x_i)+g_i(t,x_1,\ldots,x_{i-1}),~~i=2,3,4
\end{array}
\]
with $|g_i(t,x_1,\ldots,x_{i-1})|\leq \sum_{j=1}^{i-1} \gamma_{i,j}|x_j|,~i=2,3,4$ for some non-negative constants $\gamma_{i,j}$, and the origin of each (isolated) subsystem $\dot{x}_i=f_i(t,x_i)$ being exponentially stable due to the existence of a positive function $V_i(x_i)$ such that $\frac{d V_i}{dx_i}f_i(t,x_i)\leq -\alpha_i |x_i|^2$ and $|\frac{d V_i}{dx_i}|\leq \beta_i |x_i|$ for some positive constants $\alpha_i$ and $\beta_i$. Define the "interconnection" matrix as
\[
S:=\left[
\begin{array}{cccc}
\alpha_1 & 0 & 0 & 0\\
-\beta_2\gamma_{21} & \alpha_2 & 0 & 0\\
-\beta_3\gamma_{31} & -\beta_3\gamma_{32} & \alpha_3 & 0 \\
-\beta_4\gamma_{41} & -\beta_4\gamma_{42} & -\beta_4\gamma_{43} & \alpha_4 
\end{array} \right]
\]
Being lower triangular with positive diagonal terms, there exists a diagonal weight-matrix $D$ such that $DS+S^TD$ is symmetric positive definite. Then, by application of Theorem 5.4 (page 233) in \cite{khalil1996}, $x=0$ is locally exponentially 
stable.
\bibliographystyle{is-unsrt}
\bibliography{bibfile}

\end{document}